\documentclass{aastex}
\usepackage{emulateapj5,times,mathptm}
\journalinfo{The Astronomical Journal, 2003 February, astro-ph/0210475}
\slugcomment{Received 2002 August 14; accepted 2002 October 17}

\shorttitle{X-rays from the most luminous optically selected $z>4$ quasars}
\shortauthors{VIGNALI ET AL.}

\newcommand{\ltsima}{$\; \buildrel < \over \sim \;$}
\newcommand{\simlt}{\lower.5ex\hbox{\ltsima}}
\newcommand{\gtsima}{$\; \buildrel > \over \sim \;$}
\newcommand{\simgt}{\lower.5ex\hbox{\gtsima}}

\newcommand{\cgs}{ ${\rm erg~cm}^{-2}~{\rm s}^{-1}$ } 
\newcommand{\lum}{\rm erg s$^{-1}$}

\def\sun{\hbox{$\odot$}}
\def\earth{\hbox{$\oplus$}}
\def\lesssim{\mathrel{\hbox{\rlap{\hbox{\lower4pt\hbox{$\sim$}}}\hbox{$<$}}}}
\def\gtrsim{\mathrel{\hbox{\rlap{\hbox{\lower4pt\hbox{$\sim$}}}\hbox{$>$}}}}

\def\arcdeg{\hbox{$^\circ$}}
\def\arcmin{\hbox{$^\prime$}}
\def\arcsec{\hbox{$^{\prime\prime}$}}

\def\aox{$\alpha_{\rm ox}$}

\def\lumh{\rm erg s$^{-1}$ Hz$^{-1}$}
\def\ab1450{$AB_{1450(1+z)}$}

\def\asca{{\it ASCA\/}}
\def\chandra{{\it Chandra\/}}
\def\conx{{\it Constellation-X\/}}

\def\heao1{{\it HEAO-1\/}}

\def\rosat{{\it ROSAT\/}}

\def\xmm{{XMM-{\it Newton\/}}}

\begin{document}

\title{X-RAY LIGHTHOUSES OF THE HIGH-REDSHIFT UNIVERSE.\\
PROBING THE MOST LUMINOUS $Z>4$ PALOMAR DIGITAL SKY SURVEY QUASARS WITH {\it CHANDRA}}

\author{
C. Vignali,\altaffilmark{1} 
W.~N. Brandt,\altaffilmark{1} 
D.~P. Schneider,\altaffilmark{1} 
G.~P. Garmire,\altaffilmark{1}
and S. Kaspi\altaffilmark{2} 
}

\altaffiltext{1}{Department of Astronomy \& Astrophysics, The Pennsylvania State University, 
525 Davey Laboratory, University Park, PA 16802 \\
({\tt chris@astro.psu.edu, niel@astro.psu.edu, dps@astro.psu.edu, and garmire@astro.psu.edu}).}
\altaffiltext{2}{School of Physics and Astronomy, Raymond and Beverly Sackler Faculty of 
Exact Sciences, Tel-Aviv University, Tel-Aviv 69978, Israel \\
({\tt shai@wise.tau.ac.il}).}

%\date{\today}

\begin{abstract}

We present the results from exploratory \chandra\ observations of nine high-redshift \hbox{($z$=4.09--4.51)} 
optically selected quasars. These quasars, taken from the Palomar Digital Sky Survey (DPOSS), 
are among the optically brightest and most luminous $z>4$ quasars known ($M_{\rm B}$$\approx$$-$28.4 to $-$30.2). 
All have been detected by \chandra\ in exposure times of $\approx$~5--6~ks, 
tripling the number of highly luminous quasars ($M_{\rm B}<-28.4$) with X-ray detections at $z>4$. 
These quasars' average broad-band spectral energy distributions 
are characterized by steeper (more negative) 
$\alpha_{\rm ox}$ values ($\langle\alpha_{\rm ox}\rangle$=$-$1.81$\pm{0.03}$) than those of lower-luminosity, 
lower-redshift samples of quasars. 
We confirm the presence of a significant correlation between the ultraviolet magnitude and 
soft X-ray flux previously found for $z>4$ quasars. 
The joint \hbox{$\approx$~2--30~keV} rest-frame X-ray spectrum of the nine quasars 
is well parameterized by a simple power-law model whose photon index, 
$\Gamma\approx$~2.0$\pm{0.2}$, is consistent with those of lower-redshift quasars. 
No evidence for significant amounts of intrinsic absorption has been found 
($N_{\rm H}\simlt8.8\times10^{21}$~cm$^{-2}$ at 90\% confidence). 
In general, our results show that $z\approx$~4.1--4.5 quasars and local quasars 
have reasonably similar \hbox{X-ray} and broad-band spectra (once luminosity effects are 
taken into account), suggesting that the accretion mechanisms in these objects are similar. 
We also present near-simultaneous optical spectra 
for these quasars obtained with the Hobby-Eberly Telescope; 
this is the first time optical spectra have been published for seven of these objects. 
The objects presented in this paper are among the best $z>4$ targets for X-ray spectroscopy 
with \xmm\ and next-generation large-area X-ray telescopes. These will detect or constrain 
iron K$\alpha$ emission lines down to rest-frame equivalent widths of $\approx$~50~eV and intrinsic column densities down to 
$N_{\rm H}\approx$~a few $\times10^{21}$~cm$^{-2}$ at z$\approx$~4. 
We also present 45 new \rosat\ upper limits for $z\ge4$ quasars and a likely (3$\sigma$) HRI detection 
of the blazar GB~1713$+$2148 at $z=4.01$. 
\end{abstract}


\section{Introduction}

At $z\approx$~0--2, X-ray emission is thought to be a universal property of 
quasars, and X-rays have also been studied from many $z\approx$~2--4 quasars. 
In contrast to radio (e.g., Schmidt et al. 1995; Stern et al. 2000; Carilli 
et al. 2001a), 
millimeter (e.g., Omont et al. 1996, 2001; Carilli et al. 2001b), 
sub-millimeter (e.g., McMahon et al. 1999; Priddey \& McMahon 2001; Isaak et al. 2002), 
and optical (e.g., Schneider, Schmidt, \& Gunn 1989; Kennefick et al. 1995; 
Fan et al. 2001; Constantin et al. 2002) 
wavelengths, where many studies of $z>4$ quasars have been conducted, 
observations in the X-ray regime are limited but improving rapidly 
(see Brandt et al. 2002b for a review). 
Prior to 2000, there were only six published $z>4$ X-ray detections. 
These were for a 
heterogeneous mixture of objects not suitable for statistical studies because of their limited 
number and varying selection criteria. 
In the last two years, the number of $z>4$ X-ray detected quasars has increased 
to $\approx$~40 (e.g., Kaspi, Brandt, \& Schneider 2000; 
Vignali et al. 2001, hereafter V01; Brandt et al. 2002a,b; 
Bechtold et al. 2002).\footnote{See http://www.astro.psu.edu/users/niel/papers/highz-xray-detected.dat for a 
regularly updated compilation of X-ray detections at $z>4$.}

If $z>4$ quasars are radiating close to their maximum (Eddington) luminosities, 
their black holes require the assembly of a mass \hbox{$\simgt$~$10^8$--$10^9$~M$_\odot$} 
(e.g., Haiman \& Cen 2002) 
in a relativistic potential when the Universe was only $\approx$~10\% of its present age. 
X-ray studies of high-redshift quasars represent a powerful, direct way of revealing the 
physical conditions in the immediate vicinities of the first massive black holes to form in the 
Universe. 
The comparison of their \hbox{X-ray} properties with those of local quasars may 
shed light on the physical processes by 
which quasar central power sources evolve over cosmic time. 
If these quasars are indeed characterized by high accretion rates, as suggested by some models 
(e.g., Kauffmann \& Haehnelt 2000), they could plausibly 
provide evidence for some energetically important phenomena 
such as accretion-disk instabilities (e.g., Lightman \& Eardley 1974; 
Shakura \& Sunyaev 1976; Gammie 1998) and ``trapping radius'' effects 
(e.g., Begelman 1978; Rees 1978). 

The comparison of high-redshift quasar \hbox{optical-to-X-ray} spectral energy distributions (SEDs; 
quantified by means of $\alpha_{\rm ox}$, the slope of a nominal power law connecting 
2500~\AA\ and 2~keV in the rest frame) with those of lower-redshift samples 
suggests a steepening of the continuum shapes at high redshifts (V01; Bechtold et al. 2002). 
However, some studies indicate that the results may 
depend on the sample selection criteria and on the choice of the X-ray data used in the analysis 
(e.g., Kaspi et al. 2000; Brandt et al. 2002b). 
Further observations of homogeneous samples of quasars are clearly required, 
especially at the highest redshifts. 
Moreover, by determining the typical X-ray fluxes and \aox\ values of high-redshift quasars one 
should be able to assess more reliably the fraction of ionizing flux 
provided by quasars at the epoch of re-ionization 
(e.g., Becker et al. 2001; Djorgovski et al. 2001; Fan et al. 2002; Pentericci et al. 2002), 
although X-rays alone are not expected to produce a fully ionized 
intergalactic medium (e.g., Donahue \& Shull 1987; Venkatesan, Giroux, \& Shull 2001).

In this paper we present the X-ray properties of a sample of nine 
optically bright $z>4$ quasars observed with the {\it Chandra X-ray Observatory} 
during Cycle~3. 
These quasars were targeted because they are among the most optically luminous objects in the Universe 
[see Fig.~1 for a comparison with the known $z>4$ quasars; note that in the cosmology adopted in this paper, 
the bright nearby ($z=0.158$) quasar 3C~273 would have an absolute $B$-band magnitude of $\approx-26.4$]. 
%
%%%%%%%%%%%%%%
%%%	FIGURE 1: Mb vs. Redshift for Z>4 Quasars
%%%%%%%%%%%%%%%%%%
%\begin{figure}
\figurenum{1}
%\centerline{\includegraphics[angle=0,width=\textwidth]{pap_mbz_bw.eps}}
\centerline{\includegraphics[angle=0,width=8.5cm]{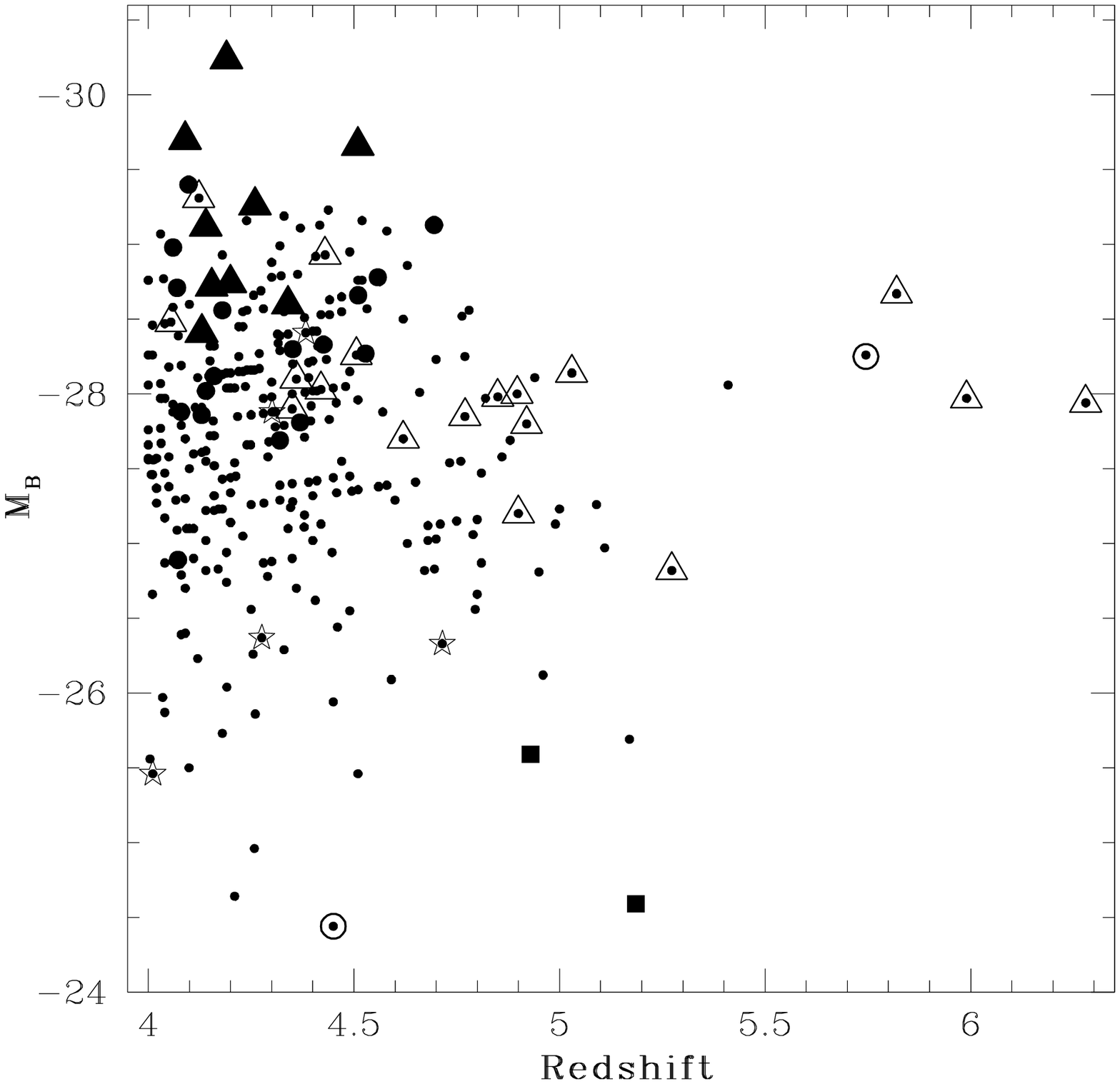}}
\figcaption{\footnotesize
Absolute $B$-band magnitude versus redshift for the known $z\ge4$ quasars 
(i.e., with $M_{\rm B}<-23$). 
Quasars with an \hbox{X-ray} detection or tight upper limit are marked 
with larger symbols. 
The PSS quasars presented in this paper are plotted as filled triangles, while open triangles indicate 
quasars observed by \chandra\ and published in previous works (e.g., V01; Brandt et al. 2002a). 
The two filled squares indicate $z>4$ quasars serendipitously discovered 
by \chandra\ at $z=4.93$ (Silverman et al. 2002) 
and $z=5.18$ (Barger et al. 2002). 
\rosat-observed quasars with tight constraints in the Kaspi et al. (2000) sample (see also V01) and 
in Appendix~A are plotted as filled circles, while 
a BALQSO at $z=5.74$ (Brandt et al. 2001b) and a $z=4.45$ X-ray selected radio-quiet quasar in the Lockman Hole 
(Schneider et al. 1998) are plotted as open circles. 
Blazars (e.g., Kaspi et al. 2000; one of the five objects is published here for the first time; 
see $\S$3.1 and Appendix~A for details) are shown as open stars. 
The uncertainties on $M_{\rm B}$ (typically $\approx$~0.4 magnitudes) are dominated by systematic errors 
due to the uncertainties in the optical magnitude measurements, the assumed optical continuum slope, and quasar variability. 
\label{fig1}}
\centerline{}
\centerline{}
%\end{figure}
%%%%%%%%%%%%%%%%%%
%%%	END of FIG.1
%%%%%%%%%%%%%%%%%%%%%%%%%%%%%%%%%%%%%%%%%%%%%%%%%%%%%%%%%%%%%%%%%%%%%%%%%%%
%
The choice of these quasars was also motivated by the possibility of obtaining 
near-simultaneous optical observations (useful to constrain their SEDs) 
with the Hobby-Eberly Telescope (HET). 
Although the number of known $z>4$ quasars has dramatically increased in the past few years 
(e.g., Djorgovski et al. 1998; Anderson et al. 2001; Peroux et al. 2001; Storrie-Lombardi et al. 2001), 
it is unlikely that large numbers of new $z>4$ quasars will be found 
with optical luminosities larger than those of the quasars presented here.  
For example, the most luminous $z>4$ quasar discovered by the Sloan Digital Sky Survey (SDSS; York et al. 2000) 
to date is SDSS~0244$-$0816 with an absolute $B$-band magnitude of $\approx$~$-$29.1. 
The quasars presented in this paper ($M_{\rm B}\approx-28.4$ to $-$30.2) are thus likely to comprise a significant fraction of 
the most luminous $z>4$ optically selected objects. 
Since optical magnitude is correlated with the soft \hbox{(0.5--2~keV)} X-ray flux 
(see, e.g., Fig.~4 of V01), these objects are ideal targets for snapshot observations 
with \chandra. 
They are probably also among the best $z>4$ quasars to select for future \hbox{X-ray} spectroscopic observations with \xmm; 
at present there is only a handful of $z>4$ quasars known to be sufficiently \hbox{X-ray} bright for 
effective \xmm\ spectroscopy. 
Our targets have been selected from the 
Palomar Digital Sky Survey (DPOSS; e.g., Djorgovski et al. 1998; hereafter these objects will be 
referred to as PSS),\footnote{See http://www.astro.caltech.edu/$\sim$george/z4.qsos 
for a listing of high-redshift quasars.} 
which covers the entire northern sky ($\delta>-3$\degr) and can find quasars up to $z\approx$~4.6 
(PSS~1347$+$4956 at $z=4.51$ is the highest redshift quasar discovered by the DPOSS thus far). 
They were identified via their distinctive red colors and confirmed with optical spectroscopy. 
We also present HET optical photometric and spectroscopic observations 
for all of the quasars in our sample. 
These were obtained nearly simultaneously with the \chandra\ observations, 
in order to provide constraints on these quasars' SEDs and moderate-quality optical spectra. 
This is the first time optical spectra have been published for seven of these 
objects.\footnote{High-quality 
optical spectra of PSS~0133$+$0400 and PSS~0209$+$0517 have been published by Peroux et al. (2001).} 

Throughout this paper we adopt $H_{0}$=70 km s$^{-1}$ Mpc$^{-1}$ in a $\Lambda$-cosmology 
with $\Omega_{\rm M}$=0.3 and $\Omega_{\Lambda}$=0.7 (e.g., Lineweaver 2001). 
For comparison with the values reported in V01 (with $H_{0}$=70 km s$^{-1}$ Mpc$^{-1}$ 
and $\Omega_{\rm M}$=1), the $\Lambda$-cosmology increases the luminosities typically 
by a factor of \hbox{$\approx$~2.3--2.4} in the redshift range 4--5.

\section{Observations and data reduction}

\subsection{\chandra\ observations and data reduction}

The nine $z>4$ optically selected quasars 
have been observed by \chandra\ during Cycle~3. 
The observation log is reported in Table~1. 
All of the sources were observed with the Advanced 
CCD Imaging Spectrometer (ACIS; Garmire et al. 2002) 
with the S3 CCD at the aimpoint. 
Faint mode was used for the event telemetry format, 
and \asca\ grade 0, 2, 3, 4 and 6 events were used in the analysis. 
No background flares occurred during the observations, 
the background being constant to within \hbox{10--20\%} in all observations. 
The average \hbox{0.5--2~keV} background value is $\approx7.6\times10^{-8}$ counts s$^{-1}$ arcsec$^{-2}$. 

Source detection was carried out with {\sc wavdetect} (Dobrzycki et al. 1999; 
Freeman et al. 2002). 
For each image, we calculated wavelet transforms (using a Mexican hat kernel) 
with wavelet scale sizes of 1, 1.4, 2, 2.8, 4, 5.7, 8, 11.3 and 16 pixels. 
Those peaks whose probability of being false were less than the threshold 
of 10$^{-6}$ were declared real; 
detections were typically achieved for the smaller wavelet scales of 1.4 pixels or less 
as expected for these distant sources. 
Due to the excellent angular resolution of \chandra\ (with a point spread function FWHM$<1$\arcsec) for on-axis positions, 
we were able to locate the X-ray sources very accurately (see Table~1), thus avoiding possible 
mis-identifications. \hbox{X-ray} positions have been determined using {\sc Sextractor} (Bertin \& Arnouts 1996) 
on the \hbox{0.5--2~keV} images and are consistent with those obtained from {\sc wavdetect}. 
Source searching was performed in four energy ranges: the ultrasoft 
band (0.3--0.5~keV), the soft band \hbox{(0.5--2~keV)}, the hard band \hbox{(2--8~keV)}, 
and the full band \hbox{(0.5--8~keV)}; 
at the average redshift of $z\approx4.2$, these energy bands correspond to the 
\hbox{1.6--2.6}, \hbox{2.6--10.4}, \hbox{10.4--41.6}, and \hbox{2.6--41.6}~keV rest-frame bands, respectively. 
%
%%%%%%%%%%%%%%%%%%%%%%%%%%%%%%%%%%%%%%%%%%%%%%%%%%%%%%%%%%%%%%%%%%%%%%%%%%%
%%%	TABLE 1: Chandra Observation Log
%%%%%%%%%%%%%%%%%%
\begin{table*}[t]
\footnotesize
\caption{\chandra\ and HET Observation Log}
\begin{center}
\begin{tabular}{lcccccccc}
\hline
\hline
Object & & & Optical & Optical & $\Delta_{\rm Opt-X}$$^{\ \rm a}$ & 
\chandra & \chandra$^{\ \rm b}$ & HET \\ 
Name & $z$ & $R$ & $\alpha_{2000}$ & $\delta_{2000}$ & (arcsec) & 
Obs. Date & Exp. (ks) & Obs. Date \\
\hline
PSS~0121$+$0347 & 4.13 & 18.3 & 01 21 26.1 & $+$03 47 07 & 1.0 & 2002 Feb 07 & 5.69 & 2001 Nov 19 \\ 
PSS~0133$+$0400 & 4.15 & 18.0 & 01 33 40.3 & $+$04 01 00 & 0.4 & 2001 Nov 26 & 6.08 & 2001 Nov 25 \\ 
PSS~0209$+$0517 & 4.14 & 17.6 & 02 09 44.6 & $+$05 17 13 & 0.1 & 2002 Jan 14 & 5.78 & 2001 Dec 09 \\ 
PSS~0926$+$3055 & 4.19 & 16.5 & 09 26 36.3 & $+$30 55 05 & 0.7 & 2002 Mar 11 & 5.29 & 2001 Dec 09 \\ 
PSS~0955$+$5940 & 4.34 & 18.3 & 09 55 11.3 & $+$59 40 31 & 0.4 & 2002 Apr 14 & 5.68 & 2001 Dec 19 \\ 
PSS~0957$+$3308 & 4.20 & 18.0 & 09 57 44.5 & $+$33 08 22 & 0.6 & 2001 Dec 24 & 6.04 & 2001 Dec 10/14 \\ 
PSS~1326$+$0743 & 4.09 & 16.9 & 13 26 11.9 & $+$07 43 58 & 0.7 & 2002 Jan 10 & 5.89 & 2001 Dec 19 \\ 
PSS~1347$+$4956 & 4.51 & 17.4 & 13 47 43.3 & $+$49 56 21 & 0.3 & 2002 May 02 & 5.89 & 2002 Feb 07 \\ 
PSS~1443$+$5856 & 4.26 & 17.5 & 14 43 40.7 & $+$58 56 53 & 0.3 & 2002 Jun 18 & 5.79 & 2002 Mar 14 \\ 
\hline
\end{tabular}
\vskip 2pt
\parbox{5.5in}
{\small\baselineskip 9pt
\footnotesize
\indent
{\sc Note. ---} 
The optical positions of the quasars have been 
derived using the source detection package {\sc Sextractor} (Bertin \& Arnouts 1996) on the 
the Palomar Digital Sky Survey (DPOSS) images. 
An average of the redshifts of the emission lines in HET spectra (excluding Ly$\alpha$; see Table~4) 
was used to determine the redshifts of the quasars. 
Only two quasars have published redshifts: PSS~0133$+$0400 ($z=4.154$; Peroux et al. 2001) and 
PSS~0209$+$0517 ($z=4.174$; Peroux et al. 2001). 
For the latter quasar, we obtain a slightly different redshift because we averaged two emission lines 
(excluding Ly$\alpha$), while Peroux et al. (2001) used Ly$\alpha$ only. 
$R$-band magnitudes have been obtained from HET photometric observations. \\
$^{\rm a}$ Distance between the optical and X-ray positions. \\
$^{\rm b}$ The \chandra\ exposure time is corrected for detector dead time. \\
}
\end{center}
\vglue-0.9cm
\end{table*}
\normalsize
%%%%%%%%%%%%%%%%%%%%%
%%%	End of Table 1
%%%%%%%%%%%%%%%%%%%%%%%%%%%%%%%%%%%%%%%%%%%%%%%%%%%%%%%%%%%%%%%%%%
%%
%%%%%%%%%%%%%%%%%%%%%%%%%%%%%%%%%%%%%%%%%%%%%%%%%%%%%%%%%%%%%%%%%%%%%%%%%%%
%%%	TABLE 2: X-ray photometry
%%%%%%%%%%%%%%%%%%
\begin{table*}[b]
\footnotesize
\caption{X-ray Counts, Hardness Ratios, Band Ratios, and Effective Photon Indices}
\begin{center}
\begin{tabular}{lccccccc}
\hline
\hline
 & \multicolumn{4}{c}{X-ray Counts$^{\rm a}$} & &  \\
\cline{2-5} \\
Object & [0.3--0.5~keV] & [0.5--2~keV] & [2--8~keV] & [0.5--8~keV] 
& Hardness Ratio$^{\rm b}$ & Band Ratio$^{\rm b}$ & $\Gamma$$^{\rm b}$ \\
\hline
PSS~0121$+$0347 & {\phn}2.0$^{+2.7}_{-1.3}$ & 58.8$^{+8.7}_{-7.6}$ & {\phn}9.8$^{+4.3}_{-3.1}$ & 68.5$^{+9.3}_{-8.3}$ & 
$-0.71^{+0.15}_{-0.13}$ & 0.17$^{+0.08}_{-0.06}$ & 2.1$^{+0.4}_{-0.3}$ \\
PSS~0133$+$0400 & {\phn}3.0$^{+2.9}_{-1.6}$ & 31.9$^{+6.7}_{-5.6}$ & {\phn}3.9$^{+3.2}_{-1.9}$ & 36.8$^{+7.1}_{-6.0}$ & 
$-0.78^{+0.24}_{-0.20}$ & 0.12$^{+0.10}_{-0.06}$ & 2.4$^{+0.6}_{-0.5}$ \\
PSS~0209$+$0517 & $<4.7$ & 21.0$^{+5.7}_{-4.5}$ & $<3.0$ & 22.9$^{+5.9}_{-4.8}$ & 
$<-0.75$ & $<0.14$ & $>2.3$ \\
PSS~0926$+$3055 & {\phn}7.0$^{+3.8}_{-2.6}$ & 45.9$^{+7.8}_{-6.8}$ & 13.9$^{+4.8}_{-3.7}$ & 59.8$^{+8.8}_{-7.7}$ & 
$-0.54^{+0.14}_{-0.12}$ & 0.30$^{+0.12}_{-0.09}$ & 1.6$\pm{0.3}$ \\
PSS~0955$+$5940 & {\phn}5.0$^{+3.4}_{-2.2}$ & {\phn}9.0$^{+4.1}_{-2.9}$ & $<6.4$ & 10.9$^{+4.4}_{-3.3}$ & 
$<-0.17$ & $<0.71$ & $>0.8$ \\
PSS~0957$+$3308 & $<3.0$ & 18.0$^{+5.3}_{-4.2}$ & $<3.0$ & 17.9$^{+5.3}_{-4.2}$ & 
$<-0.71$ & $<0.17$ & $>2.0$ \\
PSS~1326$+$0743 & {\phn}8.0$^{+4.0}_{-2.8}$ & 44.9$^{+7.8}_{-6.7}$ & 14.8$^{+4.9}_{-3.9}$ & 60.6$^{+8.8}_{-7.8}$ & 
$-0.50^{+0.13}_{-0.11}$ & 0.33$^{+0.12}_{-0.10}$ & 1.5$\pm{0.3}$ \\
PSS~1347$+$4956 & $<6.4$ & 27.0$^{+6.3}_{-5.2}$ & {\phn}4.0$^{+3.2}_{-1.9}$ & 30.9$^{+6.6}_{-5.5}$ & 
$-0.74^{+0.25}_{-0.20}$ & 0.14$^{+0.12}_{-0.08}$ & 2.2$^{+0.7}_{-0.5}$ \\
PSS~1443$+$5856 & $<3.0$ & {\phn}7.0$^{+3.8}_{-2.6}$ & $<6.4$ & 10.0$^{+4.3}_{-3.1}$ & 
$<-0.04$ & $<0.91$ & $>0.6$  \\ 
\hline
\end{tabular}
\vskip 2pt
\parbox{5.7in}
{\small\baselineskip 9pt
\footnotesize
\indent
$^{\rm a}$ Errors on the X-ray counts were computed according to Tables~1 and 2 of Gehrels (1986) 
and correspond to the 1$\sigma$ level; these were calculated assuming Poisson statistics. 
The upper limits are at the 95\% confidence level and were computed 
according to Kraft, Burrows, \& Nousek (1991). \\
$^{\rm b}$ Errors on the hardness ratios [defined as ($H-S$)/($H+S$), where $S$ is the soft-band (0.5--2~keV) 
counts and $H$ is the hard-band (2--8~keV) counts], the 
band ratios (i.e., the ratio of the \hbox{2--8~keV} to \hbox{0.5--2~keV} counts), and the 
effective photon indices are at the $\approx1\sigma$ level and have been computed 
following the ``numerical method'' described in $\S$~1.7.3 of Lyons (1991); 
this avoids the failure of the standard approximate variance formula when the number of 
counts is small (see $\S$~2.4.5 of Eadie et al. 1971). 
}
\end{center}
\vglue-0.9cm
\end{table*}
\normalsize
%%%%%%%%%%%%%%%%%%%%%
%%%	End of Table 2
%%%%%%%%%%%%%%%%%%%%%%%%%%%%%%%%%%%%%%%%%%%%%%%%%%%%%%%%%%%%%%%%%%%%%%%%%%%

All of the quasars have been detected with exposure times of \hbox{5.3--6.1~ks}. 
The {\sc wavdetect} photometry measurements are shown in Table~2; 
we have checked these results with manual aperture photometry (using a 2\arcsec\ radius circular cell) 
and found good agreement between the two techniques. 
Using a larger (4\arcsec\ radius) aperture gives results consistent with those reported in Table~2 within 
the errors. 
Table~2 also reports the hardness ratios [defined as ($H-S$)/($H+S$), where $S$ is the soft-band \hbox{(0.5--2~keV)} 
counts and $H$ is the hard-band \hbox{(2--8~keV)} counts], 
the band ratios (i.e., the ratio 
of the \hbox{2--8~keV} to \hbox{0.5--2~keV} counts), and the effective 
X-ray power-law photon indices ($\Gamma$) derived from these band ratios assuming Galactic absorption. 
All of the quasars have effective photon indices consistent, within the errors, with those 
of $z\approx$~0--2 quasars in the rest-frame \hbox{2--10~keV} band ($\Gamma\approx$~1.7--2.3; e.g., 
George et al. 2000; Mineo et al. 2000; Reeves \& Turner 2000). 
To check the results derived from the band ratios, we also performed rough spectral analyses for all of the 
sources using the Cash statistic (Cash 1979) with {\sc XSPEC} (Version 11.2.0; Arnaud et al. 1996) 
in the \hbox{0.5--8~keV} band. 
Although the number of counts for each source is too low to provide tight constraints and the signal-to-noise 
ratio dramatically decreases at energies above 2~keV, 
we found general agreement in the power-law photon indices derived with this method and those 
obtained from band ratios. 
The inclusion of the \hbox{0.3--0.5~keV} counts provides for some sources slightly different 
photon indices. However, it must be noted that \chandra\ ACIS suffered from 
calibration uncertainties in the ultrasoft band (G. Chartas 2002, private communication) 
even before the 2002 quantum efficiency loss discovery; 
therefore possible differences must be taken with caution. 

The soft-band images of the nine quasars, adaptively smoothed at the 
2$\sigma$ level using the algorithm of Ebeling, White, \& Rangarajan (2002), are shown in Fig.~2. 
The crosses mark the optical positions of the quasars. The differences between the optical and \hbox{X-ray} positions of the 
quasars are generally $\le1$\arcsec\ (see Table~1). 
%
%%%%%%%%%%%%%%%%%%%%%%%%%%%%%%%%%%%%%%%%%%%%%%%%%%%%%%%%%%%%%%%%%%%%%%%%%%%%%%%%%%%%%%%%%%%%%
%%%	FIGURE 2: Smoothed Chandra images for the 10 PSS/BRI Quasars
%%%%%%%%%%%%%%%%%%
\begin{figure*}[!ht]
\figurenum{2}
%\centerline{\includegraphics[angle=0,width=\textwidth]{newima.ps}}
\centerline{\includegraphics[angle=0,width=\textwidth]{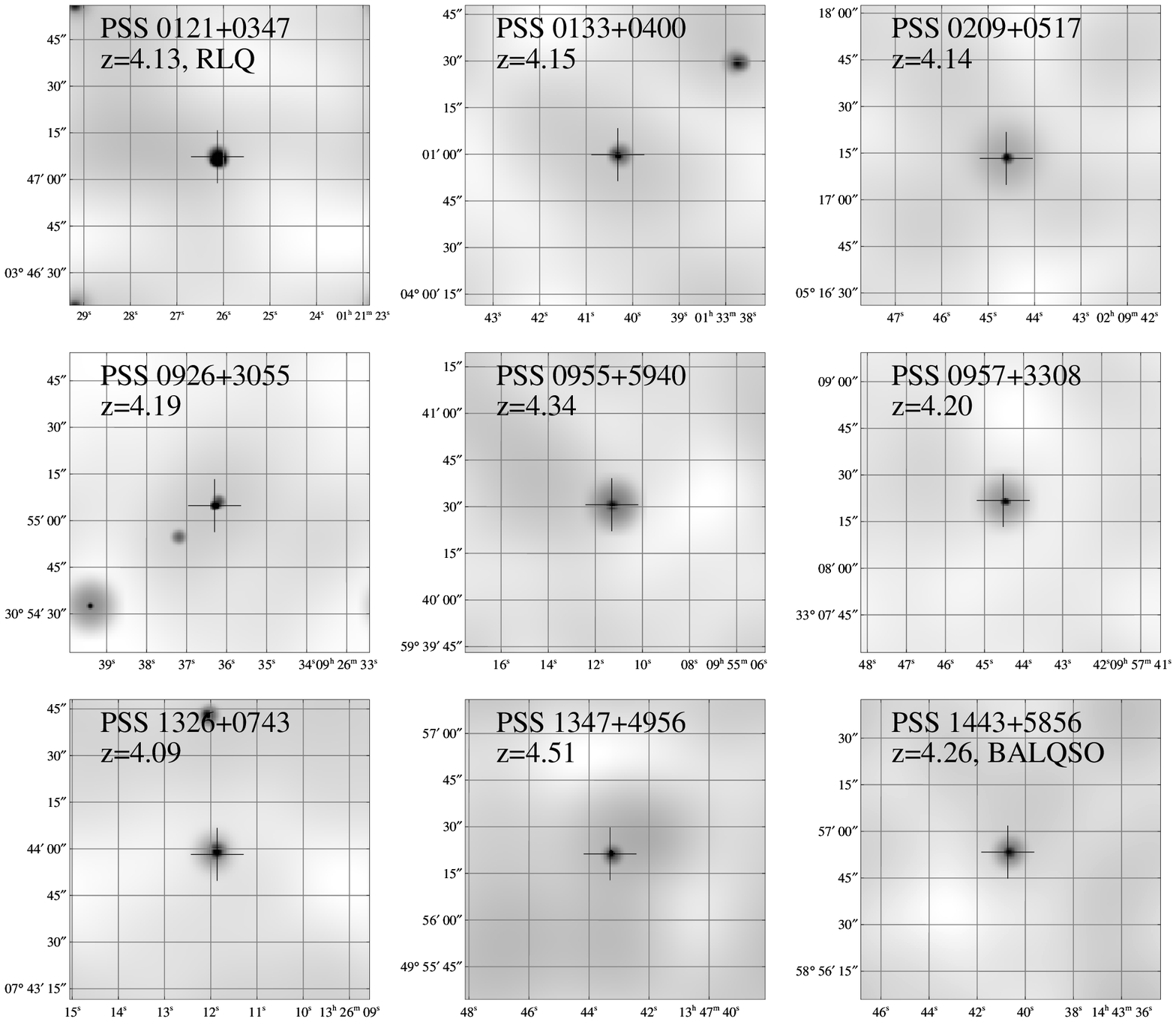}}
\vskip -6.9cm\figcaption{\footnotesize
\chandra\ soft-band (0.5--2~keV) images of the nine high-redshift PSS quasars presented in this paper. 
In each panel, the horizontal axis shows the Right Ascension, and the vertical axis shows the Declination (both in J2000 coordinates). 
Each image is 97\arcsec $\times$ 97\arcsec. 
The images have been adaptively smoothed at the 
2$\sigma$ level using the algorithm of Ebeling et al. (2002). 
Crosses mark the optical positions of the quasars. 
The possible extension of PSS~0926$+$3055 is discussed in $\S$3.5. 
\label{fig2}}
\vglue-0.2cm
\end{figure*}
%%%%%%%%%%%%%%%%%%
%%%	END of FIG.2
%%%%%%%%%%%%%%%%%%%%%%%%%%%%%%%%%%%%%%%%%%%%%%%%%%%%%%%%%%%%%%%%%%%%%%%%%%%

\subsection{Hobby-Eberly Telescope observations}

Optical spectroscopic and photometric observations of all nine quasars have been 
carried out with the 9-m Hobby-Eberly Telescope (Ramsey et al. 1998). 
The data were obtained with the HET's Marcario 
Low-Resolution Spectrograph (LRS; Hill et al. 1998a,b; Cobos Duenas et al. 1998) 
within at most $\approx$~22 days (in the rest frame) of the \chandra\ observations 
(see Table~1). 
High-redshift, high-luminosity quasars ($2<z<3.4$) have longer rest-frame optical variability time scales 
than low-redshift, low-luminosity quasars ($z<0.4$), 
and their variations over $\approx$~1.5 years (in the rest frame) amount to $\approx$~10\% (Kaspi 2001). 
Therefore, we do not expect our targets to have varied 
significantly (in the optical) during the time between the optical and X-ray observations. 
The spectroscopic observations were taken with the \hbox{300 line mm$^{-1}$} grating, an OG515 blocking filter, 
and a slit width of \hbox{2--3\arcsec} (depending on the seeing conditions), 
providing spectra from \hbox{5100--10200~\AA} 
at a resolving power of \hbox{$\approx$~240--300}. 
The exposure time 
was chosen to be 900~s for all but one of quasars, the exception being 
for PSS~0957$+$3308 (two close observations of 800 and 900~s, respectively). 
The spectroscopic data were reduced using 
standard {\sc IRAF} routines. Bias exposures taken on each night were stacked and subtracted, and 
bias-subtracted frames were then flat fielded in a standard manner using 
internal lamp flats obtained during the same night. 
The wavelength calibration was performed using a Ne arc lamp, 
while the flux calibration was achieved using one standard star for each night. 
The calibrated spectra are shown in Fig.~3.\footnote{The HET optical spectra are available at 
http://www.astro.psu.edu/users/niel/papers/papers.html.} 
%
%%%%%%%%%%%%%%
%%%	FIGURE 3: HET Optical Spectra for the 9 PSS Quasars
%%%%%%%%%%%%%%%%%%
\begin{figure*}[!ht]
\figurenum{3}
%\centerline{\includegraphics[angle=0,width=\textwidth]{optspec.ps}}
\centerline{\includegraphics[angle=0,width=\textwidth]{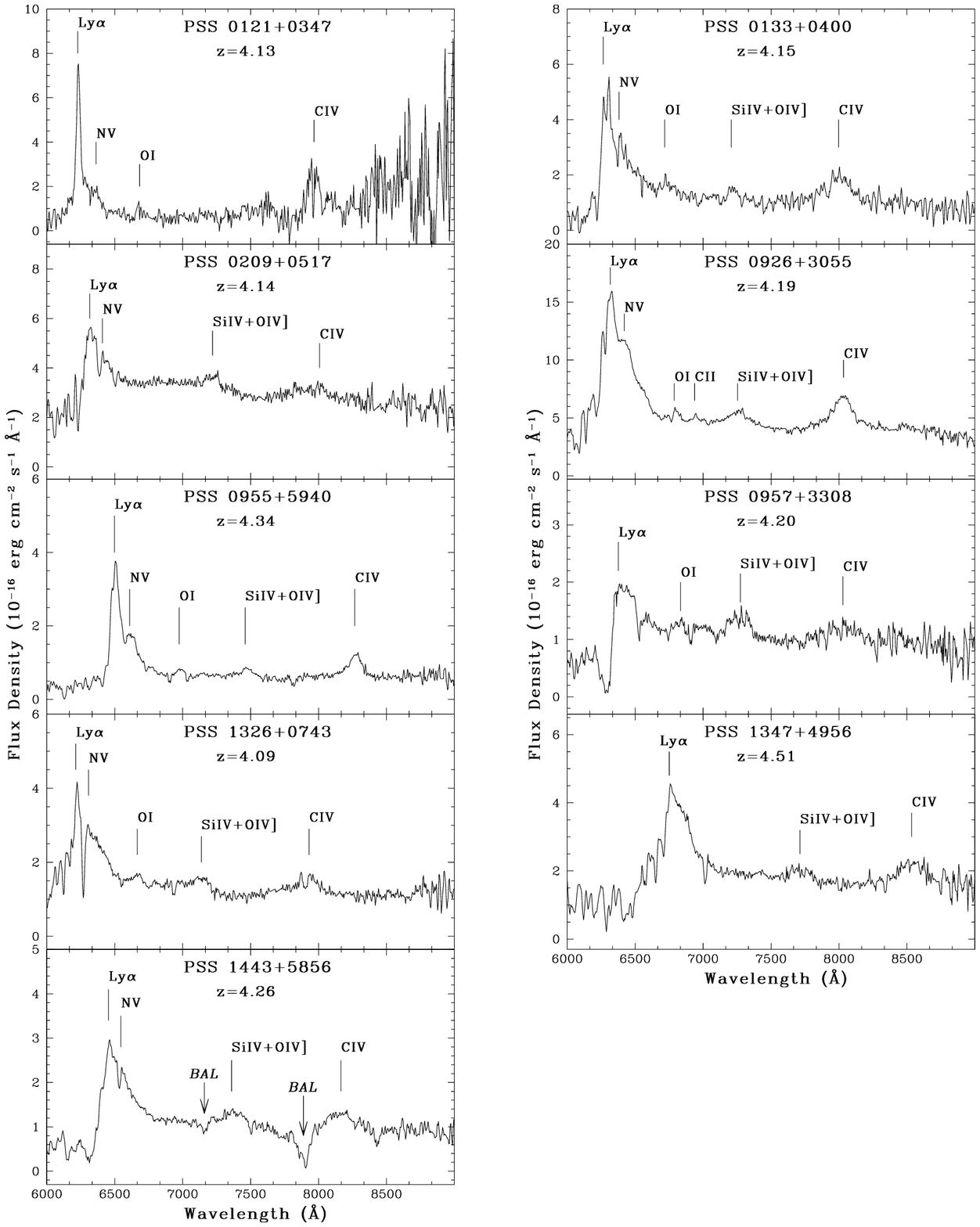}}
\vskip -3.6cm\figcaption{\footnotesize
Optical spectra of the nine PSS quasars presented in this paper obtained with the Hobby-Eberly Telescope 
at a resolving power of \hbox{$\approx$~240--300}. 
The principal emission lines are indicated. Some of them are not reported in Table~4 because of the 
uncertainties in their measurements.  
The arrows in the panel for PSS~1443$+$5856 
mark the \ion{Si}{4} and \ion{C}{4} broad absorption lines. 
\label{fig3}}
\vglue-0.2cm
\end{figure*}
%%%%%%%%%%%%%%%%%%
%%%	END of FIG.3
%%%%%%%%%%%%%%%%%%%%%%%%%%%%%%%%%%%%%%%%%%%%%%%%%%%%%%%%%%%%%%%%%%%%%%%%%%%
%
Given that there are flux losses due to the slit size and that we did not demand 
photometric conditions for our optical observations, there is considerable uncertainty in the flux density normalization. 

To address the long-term optical variability of the present sample of quasars, 
we also obtained 120~s $R$-band exposures immediately before the spectroscopic observations. 
Each image covered a field $\approx4\arcmin$ on a side, and using United State Naval Observatory (USNO) 
photometry of DPOSS objects 
it was possible to obtain an approximate photometric calibration. 
Two objects (PSS~0926$+$3055 and PSS~0955$+$5940) showed significant variations (see $\S$4).

\section{X-ray analysis}

The principal X-ray, optical, and radio properties of our targets are given in Table~3. 
A description is as follows: \\
{\sl Column (1)}. --- The name of the source. \\
{\sl Column (2)}. --- The Galactic column density (from Dickey \& Lockman 1990) 
in units of \hbox{10$^{20}$ cm$^{-2}$}. \\
{\sl Column (3)}. --- The monochromatic rest-frame \ab1450\ magnitude (defined in $\S$3b of 
Schneider et al. 1989) with estimated errors $\simlt0.1$~mag, corrected for Galactic reddening. 
\ab1450\ magnitudes have been derived from HET $R$-band magnitudes assuming the 
empirical relationship \ab1450=$R-$0.684\ $z+3.10$ (effective in the redshift range covered by our sample); 
see Table~2 of Kaspi et al. (2000). 
\ab1450\ magnitudes have not been derived directly from the spectra because of flux-calibration uncertainties. \\
{\sl Columns (4) and (5)}. --- The 2500~\AA\ rest-frame flux density and luminosity. 
These were computed from the \ab1450\ magnitude assuming an optical power-law slope of 
$\alpha$=$-$0.79 \hbox{($S_{\nu}$ $\propto$ $\nu^{\alpha}$}; Fan et al. 2001) 
to keep consistency with V01; 
see V01 for further details about the choice of the optical slope. 
Changing the power-law slope of the optical continuum from $\alpha$=$-$0.79 (Fan et al. 2001) to 
$\alpha$=$-$0.5 (e.g., Vanden Berk et al. 2001) reduces the 2500~\AA\ rest-frame flux densities and luminosities 
by $\approx$~15\%. \\
{\sl Column (6)}. --- The absolute $B$-band magnitude calculated according to the 
formula
\begin{equation}
M_{\rm B}=AB_{1450(1+z)}-5\log D_{\rm L}+5+2.5\log (1+z)-0.83
%+1.205\times\alpha_{\rm o}+0.12
\end{equation}
where $D_{\rm L}$ is the luminosity distance. 
We note that changing the power-law slope of the optical continuum from $\alpha$=$-$0.79 to 
$\alpha$=$-$0.5 gives fainter absolute $B$-band magnitudes 
by $\approx$~0.35 magnitudes. \\
{\sl Columns (7) and (8)}. --- The observed count rate in the \hbox{0.5--2~keV} band and the corresponding 
flux (corrected for Galactic absorption).  
The \hbox{0.5--2~keV} counts have been converted into fluxes (using {\sc PIMMS} Version 3.2d; Mukai 2002) 
assuming a power-law model with $\Gamma=2$ 
(as derived for samples of lower redshift quasars; e.g., George et al. 2000; Mineo et al. 2000; Reeves \& Turner 2000). 
Changes of the photon index in the range $\Gamma=$1.7--2.3 lead to only 
a few percent change in the measured X-ray flux. 
The soft X-ray flux derived from the \hbox{0.5--2~keV} counts 
has also been compared with that derived using the full-band \hbox{(0.5--8~keV)} counts. 
This comparison gave similar results to within typically 10--20\%. \\
{\sl Columns (9) and (10)}. --- The rest-frame 2~keV flux density and luminosity. These were 
computed assuming the same power-law photon index as in the count-rate-to-flux conversion. \\
{\sl Column (11)}. ---  The \hbox{2--10~keV} rest-frame luminosity corrected for Galactic absorption. \\
{\sl Column (12)}. ---  The optical-to-X-ray power-law slope, $\alpha_{\rm ox}$, defined as 
\begin{equation}
\alpha_{\rm ox}=\frac{\log(f_{\rm 2~keV}/f_{2500~\mbox{\scriptsize\AA}})}{\log(\nu_{\rm 2~keV}/\nu_{2500~\mbox{\scriptsize\AA}})}
\end{equation}
where $f_{\rm 2~keV}$ and $f_{2500~\mbox{\scriptsize\AA}}$ are the flux densities at 2~keV and 2500~\AA, 
respectively.\footnote{$f_{\rm 2~keV}$ has been computed from the observed-frame 0.5--2~keV 
flux (assuming an \hbox{X-ray} power-law photon index of $\Gamma=2$); this is
required because ACIS has limited spectral response below 0.5~keV. 
Thus, the derived \aox\ values are actually based on the relative 
amount of X-ray flux in the 0.5$(1+z)$~keV to 2$(1+z)$~keV rest-frame band.}
The reported errors have been computed at the $\approx1\sigma$ level 
following the ``numerical method'' described in $\S$~1.7.3 of Lyons (1991). 
Both the statistical uncertainties on the X-ray count rates 
and the effects of possible changes in the X-ray ($\Gamma\approx$~1.7--2.3) 
and optical \hbox{($\alpha$$\approx$~$-$0.5 to $-$0.9}; e.g., 
Vanden Berk et al. 2001; Schneider et al. 2001) continuum shapes 
have been taken into account (see $\S$3 of V01 for further details). 
We note that changing the power-law slope of the optical continuum from $\alpha$=$-$0.79 to 
$\alpha$=$-$0.5 induces a small increase ($\approx$~0.03) in the \aox\ values. \\
{\sl Column (13)}. --- The radio-loudness parameter (e.g., Kellermann et al. 1989), 
defined as \hbox{$R$=$f_{\rm 5~GHz}/f_{\rm 4400~\mbox{\scriptsize\AA}}$} (rest frame). 
The rest-frame 5~GHz flux density was computed from the FIRST (Becker, White, \& Helfand 1995) or NVSS 
(Condon et al. 1998) observed-frame 1.4~GHz flux density assuming a radio power-law slope of $\alpha=-0.8$. 
The upper limits reported in the table are at the 3\/$\sigma$ level. 
The rest-frame $f_{\rm 4400~\mbox{\scriptsize\AA}}$ flux density 
has been computed from the \ab1450\ magnitude assuming an 
optical power-law slope of $\alpha=-0.79$. 
Radio-loud quasars (RLQs) are characterized by $R>100$, whereas radio-quiet quasars (RQQs) 
have $R<10$ (e.g., Kellermann et al. 1989). 
PSS~0121$+$0347 is radio loud and shows the flattest \aox\ value in our sample, 
consistent with results from previous studies (e.g., Zamorani et al. 1981); 
the other eight objects are radio quiet.

\subsection{X-ray flux versus \ab1450\ magnitude}

\chandra\ and other exploratory X-ray observations have defined the typical fluxes and luminosities of 
$z>4$ quasars (e.g., Kaspi et al. 2000; V01; Brandt et al. 2002a, 2002b; Bechtold et al. 2002). 
Figure~4 shows the observed-frame, Galactic absorption-corrected \hbox{0.5--2~keV} 
flux versus \ab1450\ magnitude for a compilation of $z\ge4$ Active Galactic Nuclei (AGNs). 
The objects presented in this paper are shown as filled triangles, while open triangles and large downward-pointing 
arrows are previous \chandra\ observations of high-redshift quasars (e.g., V01; Brandt et al. 2002a), 
mostly taken from the SDSS. 
Circled triangles are RLQs, according to the definition in $\S3$.  
In the following analyses and plots we exclude all but one (SDSS~120441.73$-$002149.6) of 
Bechtold Cycle~2 $z>4$ quasars (Bechtold et al. 2002) 
since they were measured in a different way than all other objects presented here. 
The \rosat-detected RQQs and RLQ are plotted as filled circles and a circled star, respectively 
(Kaspi et al. 2000; V01), while small downward-pointing arrows indicate \rosat\ upper limits, most of which 
are presented here for the first time 
(see Appendix~A for further information). 
Blazars from the Kaspi et al. (2000) sample are plotted as open stars. 
One object (the $z=4.01$ blazar GB~1713$+$2148; Hook \& McMahon 1998) 
has its X-ray data published here for the first time and is a $\approx$~3$\sigma$ \rosat\ HRI detection. 
As a comparison sample, in Fig.~4 we also plot as open circles the $z=4.45$ X-ray selected RQQ RX~J1052$+$5719 
(Schneider et al. 1998), the $z=5.74$ Broad Absorption Line (BAL) quasar SDSS~1044$-$0125 (Fan et al. 
%%%%%%%%%%%%%%%%%%%%%%%%%%%%%%%%%%%%%%%%%%%%%%%%%%%%%%%%%%%%%%%%%%%%%%%%%%%
%%%	TABLE 3: Optical + X-ray properties
%%%%%%%%%%%%%%%%%%
\end{multicols}
\begin{deluxetable}{lcccccccccccc}
\rotate
\tablecolumns{13}
\tabletypesize{\scriptsize}
\tablewidth{0pt}
\tablecaption{Properties of $z>4$ Quasars Observed by \chandra}
\tablehead{ 
\colhead{Object} & \colhead{$N_{\rm H}$\tablenotemark{a}} & 
\colhead{$AB_{1450(1+z)}$} & \colhead{$f_{2500}$\tablenotemark{b}} & \colhead{$\log (\nu L_\nu )_{2500}$} & 
\colhead{$M_B$} & \colhead{Count~rate\tablenotemark{c}} & 
\colhead{$f_{\rm x}$\tablenotemark{d}} & \colhead{$f_{\rm 2\/keV}$\tablenotemark{e}} & 
\colhead{$\log (\nu L_\nu )_{\rm 2\/keV}$} & \colhead{$\log (L_{\rm 2-10~keV})$\tablenotemark{f}} & 
\colhead{$\alpha_{\rm ox}$\tablenotemark{g}} & \colhead{$R$\tablenotemark{h}} \\
\colhead{(1)} & \colhead{(2)} & \colhead{(3)} & \colhead{(4)} & \colhead{(5)} & \colhead{(6)} & \colhead{(7)} &  
\colhead{(8)} & \colhead{(9)} & \colhead{(10)} & \colhead{(11)} & \colhead{(12)} & \colhead{(13)}
}
\startdata
PSS~0121$+$0347 & 3.27 & 18.5 & {\phn}2.22 & 46.9 & $-$28.4 & 10.30$^{+1.60}_{-1.30}$ & 
34.7$^{+5.4}_{-4.4}$ & 26.4 & 45.6 & 45.8 & $-$1.51$\pm{0.06}$ & 300.2\tablenotemark{i} \\
PSS~0133$+$0400 & 3.06 & 18.2 & {\phn}2.93 & 47.1 & $-$28.7 & {\phn}5.25$^{+1.10}_{-0.92}$ & 
17.4$^{+3.7}_{-3.0}$ & 13.4 & 45.3 & 45.5 & $-$1.67$\pm{0.06}$ & $<4.4$\tablenotemark{i} \\
PSS~0209$+$0517 & 4.73 & 17.8 & {\phn}4.24 & 47.2 & $-$29.1 & {\phn}3.64$^{+0.98}_{-0.78}$ & 
12.8$^{+3.4}_{-2.8}$ & {\phn}9.8 & 45.2 & 45.4 & $-$1.78$^{+0.07}_{-0.06}$ & $<3.0$\tablenotemark{i} \\
%
% Change in AB1450(1+z): 17.2-->16.9 [before using HET R-band]
PSS~0926$+$3055 & 1.88 & 16.7 & 11.70 & 47.7 & $-$30.2 & {\phn}8.67$^{+1.47}_{-1.28}$ & 
27.7$^{+4.7}_{-4.1}$ & 21.4 & 45.5 & 45.7 & $-$1.82$\pm{0.06}$ & $<0.3$\tablenotemark{j} \\
%
% Change in AB1450(1+z): 17.6-->18.1 [before using HET R-band]
PSS~0955$+$5940 & 1.28 & 18.4 & {\phn}2.44 & 47.0 & $-$28.6 & {\phn}1.58$^{+0.72}_{-0.51}$ & 
{\phn}5.0$^{+2.2}_{-1.6}$ & {\phn}3.9 & 44.8 & 45.0 & $-$1.84$\pm{0.08}$ & $<5.4$\tablenotemark{i} \\
PSS~0957$+$3308 & 1.52 & 18.2 & {\phn}2.93 & 47.1 & $-$28.7 & {\phn}2.98$^{+0.88}_{-0.69}$ & 
{\phn}9.4$^{+2.8}_{-2.2}$ & {\phn}7.3 & 45.1 & 45.3 & $-$1.77$\pm{0.07}$ & $<1.1$\tablenotemark{j} \\
PSS~1326$+$0743 & 2.01 & 17.2 & {\phn}7.36 & 47.4 & $-$29.7 & {\phn}7.62$^{+1.33}_{-1.13}$ & 
24.4$^{+4.3}_{-3.6}$ & 18.6 & 45.4 & 45.7 & $-$1.77$\pm{0.06}$ & $<0.5$\tablenotemark{j} \\
PSS~1347$+$4956 & 1.23 & 17.4 & {\phn}6.12 & 47.4 & $-$29.7 & {\phn}4.59$^{+1.07}_{-0.89}$ & 
14.3$^{+3.4}_{-2.7}$ & 11.8 & 45.3 & 45.5 & $-$1.81$^{+0.07}_{-0.06}$ & {\phn}{\phn}0.1\tablenotemark{k} \\
PSS~1443$+$5856 & 1.50 & 17.7 & {\phn}4.64 & 47.3 & $-$29.3 & {\phn}1.21$^{+0.83}_{-0.45}$ & 
{\phn}3.8$^{+2.6}_{-1.4}$ & {\phn}3.0 & 44.7 & 44.9 & $-$1.99$^{+0.10}_{-0.09}$ & $<2.8$\tablenotemark{i} \\
\tableline
\enddata
\tablecomments{Luminosities are computed using $H_{0}$=70 km s$^{-1}$ Mpc$^{-1}$, 
$\Omega_{\rm M}$=0.3, and $\Omega_{\Lambda}$=0.7.}
\tablenotetext{a}{From Dickey \& Lockman (1990), in units of $10^{20}$ cm$^{-2}$ .}
\tablenotetext{b}{2500~\AA\ rest-frame flux density, in units of $10^{-27}$ erg cm$^{-2}$ s$^{-1}$ Hz$^{-1}$.}
\tablenotetext{c}{Observed count rate computed in the 0.5--2~keV band, in units of $10^{-3}$  counts s$^{-1}$.}
\tablenotetext{d}{Galactic absorption-corrected flux in the observed 0.5--2~keV band, in units 
of $10^{-15}$ erg cm$^{-2}$ s$^{-1}$.} 
\tablenotetext{e}{Rest-frame 2~keV flux density, in units of $10^{-32}$ erg cm$^{-2}$ s$^{-1}$ Hz$^{-1}$.} 
\tablenotetext{f}{2--10~keV rest-frame luminosity corrected for Galactic absorption, 
in units of erg s$^{-1}$.}
\tablenotetext{g}{Errors have been computed following the ``numerical method'' described in $\S$~1.7.3 of Lyons (1991); 
both the statistical uncertainties on the X-ray count rates 
and the effects of the observed ranges of the X-ray and optical continuum shapes have been taken into account 
(see text for details; see also $\S$3 in V01).} 
\tablenotetext{h}{Radio-loudness parameter, defined as 
$R$ = $f_{\rm 5~GHz}/f_{\rm 4400~\mbox{\scriptsize\AA}}$ (rest frame) (e.g., Kellermann et al. 1989). 
The rest-frame 5~GHz flux density is computed from the observed 1.4~GHz flux density 
assuming a radio power-law slope of $\alpha=-0.8$, with $f_{\nu}\propto~\nu^{\alpha}$.}
\tablenotetext{i}{1.4~GHz flux density from NVSS (Condon et al. 1998).}
\tablenotetext{j}{1.4~GHz flux density from FIRST (Becker et al. 1995).}
\tablenotetext{k}{1.4~GHz Very Large Array flux density from Carilli et al. (2001a).}
\label{tab3}
\end{deluxetable}
\begin{multicols}{2}
%%%%%%%%%%%%%%%%%%%%%
%%%	End of Table 3
%%%%%%%%%%%%%%%%%%%%%%%%%%%%%%%%%%%%%%%%%%%%%%%%%%%%%%%%%%%%%%%%%%%%%%%%%%%
%

\noindent 
2000; Brandt et al. 2001b), 
and the $z=4.42$ Seyfert galaxy VLA~J1236$+$6213 (Waddington et al. 1999; Brandt et al. 2001c). 
The three filled squares show two quasars and a Seyfert galaxy serendipitously discovered by \chandra\ 
(Silverman et al. 2002; Barger et al. 2002) at $z=4.93$, $z=5.18$, and $z=4.13$. 
The slanted lines show \aox\ spectral indices of $-$1.5 
%\noindent 
and $-$1.8 at $z=4.2$, which is 
the average redshift for the sample discussed in this paper. 

%%%%%%%%%%%%%%
%%%	FIGURE 4: Fx [0.5-2 keV, un.] versus AB1450(1+z) 
%%%%%%%%%%%%%%%%%%
%\begin{figure}
\figurenum{4}
%\centerline{\includegraphics[angle=0,width=\textwidth]{pap_fxab_bw.eps}}
\centerline{\includegraphics[angle=0,width=8.5cm]{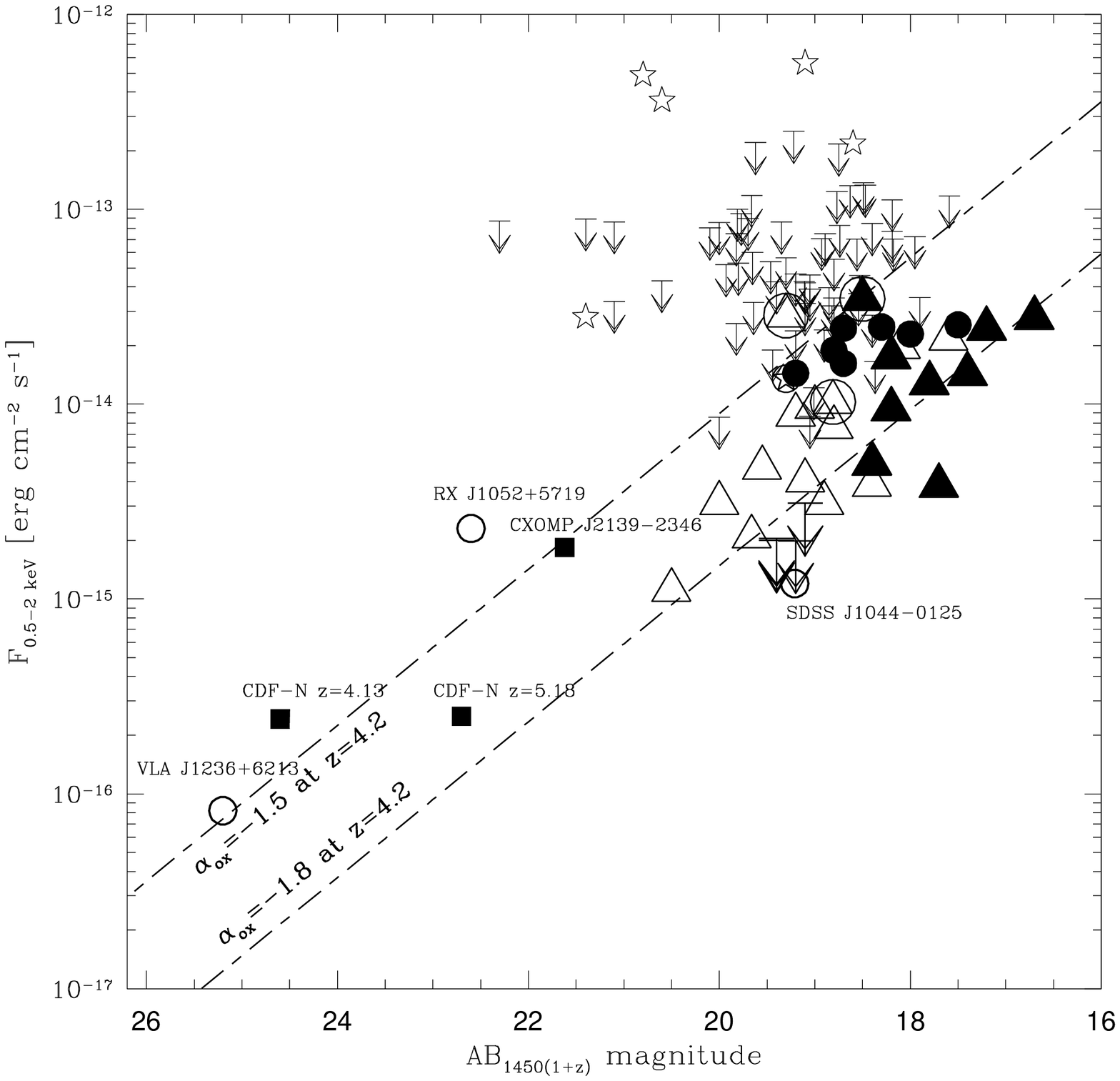}}
\figcaption{\footnotesize
Observed-frame, Galactic absorption-corrected \hbox{0.5--2~keV} flux versus $AB_{1450(1+z)}$ 
magnitude for $z\ge4$ AGNs. 
The objects presented in this paper are plotted as filled triangles, 
while the open triangles and large downward-pointing arrows indicate previous \chandra\ observations 
of high-redshift quasars (V01; Brandt et al. 2002a). Circled triangles are RLQs. 
The \rosat-detected RQQs and RLQ are plotted as filled circles and a circled star, respectively 
(Kaspi et al. 2000; V01), while 
small downward-pointing arrows indicate \rosat\ upper limits (most of the upper limits 
are presented here for the first time; see Appendix~A for details). 
Blazars from the Kaspi et al. (2000) sample are plotted as open stars (one object is published here for 
the first time; see Appendix~A). 
For comparison, we have plotted as open circles the $z=4.45$ X-ray selected RQQ RX~J1052$+$5719 
(Schneider et al. 1998), the $z=5.74$ RQQ SDSS~J1044$-$0125 (Fan et al. 2000; Brandt et al. 2001b), and the 
$z=4.42$ Seyfert galaxy VLA~J1236$+$6213 (Waddington et al. 1999; Brandt et al. 2001c). 
The three filled squares indicate $z>4$ AGNs serendipitously discovered by \chandra\ 
(Silverman et al. 2002; Barger et al. 2002). 
The slanted lines show \aox=$-$1.5 and \aox=$-$1.8 loci at $z=4.2$ (the average redshift of the present sample), 
assuming the same X-ray and optical spectral shapes used in the text. 
\label{fig4}}
%\end{figure}
\centerline{}
\centerline{}
%%%%%%%%%%%%%%%%%%
%%%	END of FIG.4
%%%%%%%%%%%%%%%%%%%%%%%%%%%%%%%%%%%%%%%%%%%%%%%%%%%%%%%%%%%%%%%%%%%%%%%%%%%
%

As shown in Fig.~4, quasars at $z>4$ are generally faint \hbox{X-ray} sources, with \hbox{0.5--2~keV} 
fluxes \hbox{$<4\times~10^{-14}$ \cgs}. 
The PSS quasars presented in this paper are among the \hbox{X-ray} brightest known and 
allow study of a region of the luminosity-redshift parameter space different 
from that of most of the SDSS quasars observed in X-rays to date (e.g., V01; Brandt et al. 2002a), 
which are usually optically less luminous and located at higher redshift. 
Given the X-ray fluxes found from \chandra\ snapshot observations, it is clear that 
PSS and BRI\footnote{BRI quasars have been selected from the B/R/I survey 
(Irwin, McMahon, \& Hazard 1991) and then confirmed spectroscopically (e.g., Storrie-Lombardi et al. 1996, 2001). 
Their selection criteria and optical magnitude range are similar to those of PSS quasars. 
See V01 and Bechtold et al. (2002) for the X-ray properties of BRI quasars in \chandra\ observations.} 
quasars represent the best known RQQs at $z>4$ for longer \xmm\ spectroscopic observations. 
Individual \xmm\ spectra, coupled with stacked \chandra\ spectra, 
can provide tighter spectral constraints and 
possibly allow searches for spectral features such as iron K$\alpha$ lines. 

%%%%%%%%%%%%%%%%%%%%%%%%%%%%%%%%%%%%%%%%%%%%%%%%%%%%%%%%%%%%%%%%%%%%%%%%%%%%%%%%%%%%%%%%%%%%%%%%%%%%%%%%%%%%%%%%%%%%%%%%%%
% Statistics as 28 June 2002: inclusion of all the Chandra QSOs without exceptions [also SDSS not plotted in these figures]
%%%%%%%%%%%%%%%%%%%%%%%%%%%%%%%%%%%%%%%%%%%%%%%%%%%%%%%%%%%%%%%%%%%%%%%%%%%%%%%%%%%%%%%%%%%%%%%%%%%%%%%%%%%%%%%%%%%%%%%%%%
In Fig.~4 there is a correlation between the soft \hbox{X-ray} flux (corrected for Galactic absorption) 
and \ab1450\ magnitude. 
Given the presence of upper limits, 
to study this correlation we have used the {\sc ASURV} software package Rev~1.2 (LaValley, Isobe, \& Feigelson 1992). 
To evaluate the significance of this correlation, we used several methods available in {\sc ASURV}, 
namely the Cox regression proportional hazard (Cox 1972), 
the generalized Kendall's $\tau$ (Brown, Hollander, \& Korwar 1974), and the Spearman's $\rho$ models. 
Using the \chandra\ RQQs, the correlation is significant at the 99.5--99.8\% and 99.99\% levels 
when all the PSS/BRI and PSS/BRI/SDSS quasars are taken into account, respectively. 
In the latter case, the best-fit relationship is parameterized by 
\hbox{$f_{\rm X}=[-(6.9\pm{1.0})\ AB_{1450(1+z)}+(137\pm{19})]\times 10^{-15}$ \cgs}. 
Note that the scatter around the best-fit relationship is significant (a factor of $\approx$~3). 
The BALQSOs have been excluded in these correlation analyses because they are known to be 
significantly absorbed in the X-ray band (e.g., Brandt, Laor, \& Wills 2000; Gallagher et al. 2001, 2002; Green et al. 2001; 
see also V01 for discussion about the BALQSOs previously observed and undetected by \chandra\ at $z>4$). 
Including also the RQQs with \rosat\ detections or tight upper limits (i.e., \aox$<-1.5$) provides a 
significance $>99.99$\% for the above correlation. 
The observed correlation indicates that the optical and \hbox{X-ray} emission are likely to be powered by the same 
engine, namely accretion onto a supermassive black hole. 
The large \hbox{2--10~keV} rest-frame luminosities of the PSS objects investigated in this paper \hbox{($\approx$~$10^{44.9-45.8}$ \lum)} 
suggest that the \hbox{X-ray} emission is mostly nuclear and not related to the starburst component whose existence has been 
suggested by recent observations at millimeter and sub-millimeter wavelengths (e.g., Omont et al. 2001; Isaak et al. 2002; 
see also Elvis, Marengo, \& Karovska 2002 for a different interpretation).

\subsection{\aox\ results}

The average \aox\ for the eight RQQs of the present PSS sample is $\langle\alpha_{\rm ox}\rangle$=$-$1.81$\pm{0.03}$ 
(the quoted errors represent the standard deviation of the mean), with 
the BALQSO PSS~1443$+$5856 (see $\S$4) being the quasar 
with the steepest \aox\ in our sample (\aox=$-1.99^{+0.10}_{-0.09}$). 
Even after removing the BALQSO, the average \aox\ is $-$1.78$\pm{0.02}$, still 
considerably steeper than that obtained for 
samples of lower-luminosity, local quasars 
[e.g., the Bright Quasar Survey (BQS; Schmidt \& Green 1983) quasars at \hbox{$z<0.5$} have 
$\langle\alpha_{\rm ox}\rangle$=$-$1.56$\pm{0.02}$; see Brandt et al. 2000 and V01]. 
The same applies when all the PSS/BRI ($\langle\alpha_{\rm ox}\rangle$=$-$1.78$\pm{0.03}$) 
or the SDSS ($\langle\alpha_{\rm ox}\rangle$=$-$1.75$\pm{0.03}$) $z>4$ quasars observed by \chandra\ are taken into account. 
These average \aox\ values are slightly flatter (typically by 0.02--0.03) when the BALQSOs are excluded from the analysis. 
One possible explanation for the more negative \aox\ values seen at $z>4$ is the presence of \hbox{X-ray} absorption; 
however, we find no evidence of \hbox{X-ray absorption} either from the individual quasars' spectra 
(see Table~2) or from the joint spectral fitting discussed in $\S$3.3. 
Another possibility is that \aox\ depends upon 2500~\AA\ rest-frame luminosity 
(e.g., Avni, Worrall, \& Morgan 1995) or redshift (e.g., Bechtold et al. 2002). 
We have searched for significant correlations of \aox\ with either 2500~\AA\ rest-frame luminosity 
(Fig.~5a) or redshift (Fig.~5b) \ at \ \ \ 

%%%%%%%%%%%%%%
%%%	FIGURE 5 [upper panel]: aox versus 2500 A Luminosity for z>4 Optically Selected RQQs
%%%	FIGURE 5 [lower panel]: aox versus Redshift for z>4 Optically Selected RQQs
%%%%%%%%%%%%%%%%%%
%\begin{figure}
\figurenum{5}
\vglue-1.8cm
%\centerline{\includegraphics[width=0.8\textwidth,angle=-90]{pap_l2500aoxz4_bw.eps}}
\centerline{\includegraphics[width=8.cm,angle=-90]{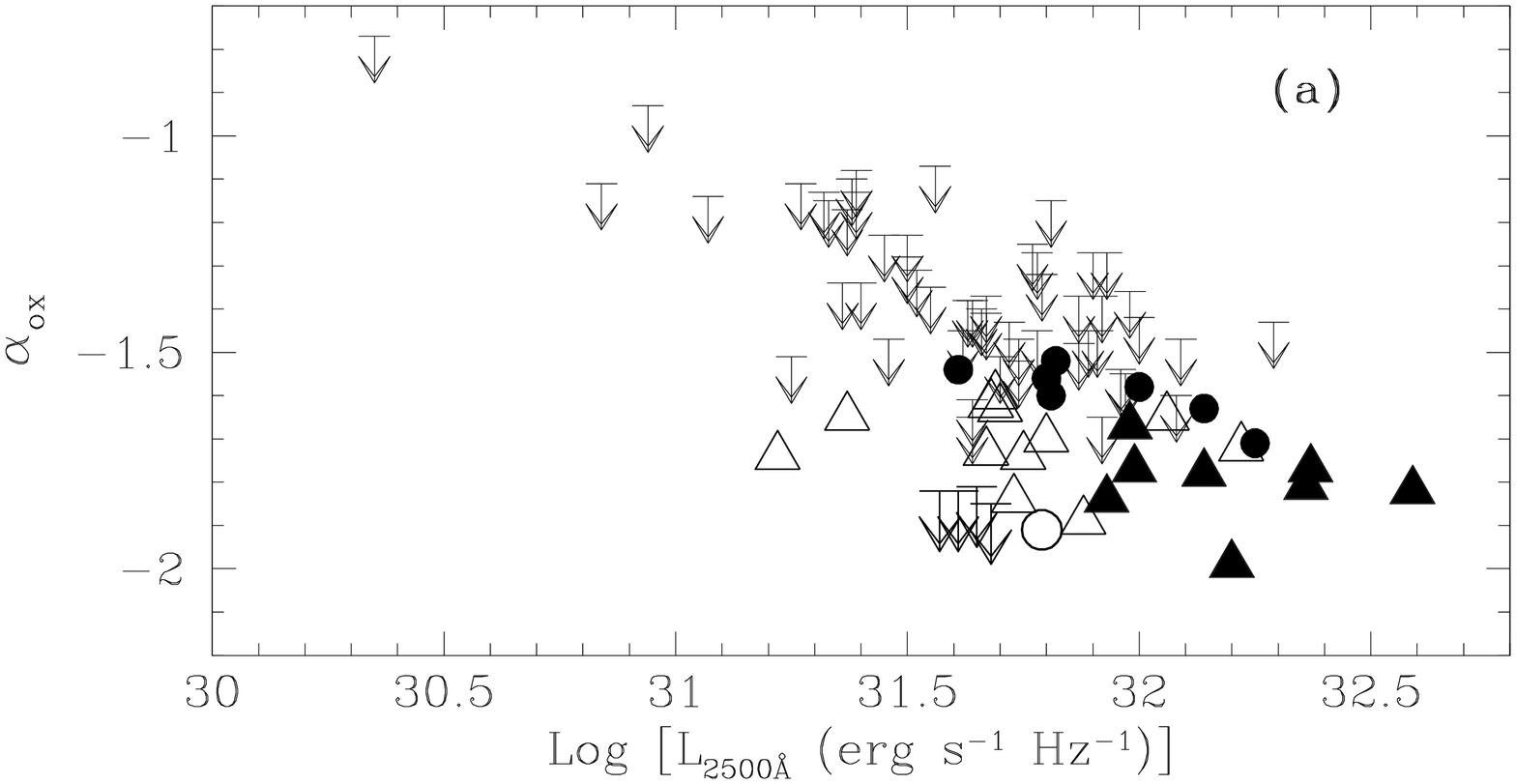}}
\vskip -2.7cm
%\centerline{\includegraphics[width=0.8\textwidth,angle=-90]{pap_aoxz4_bw.eps}}
\centerline{\includegraphics[width=8.cm,angle=-90]{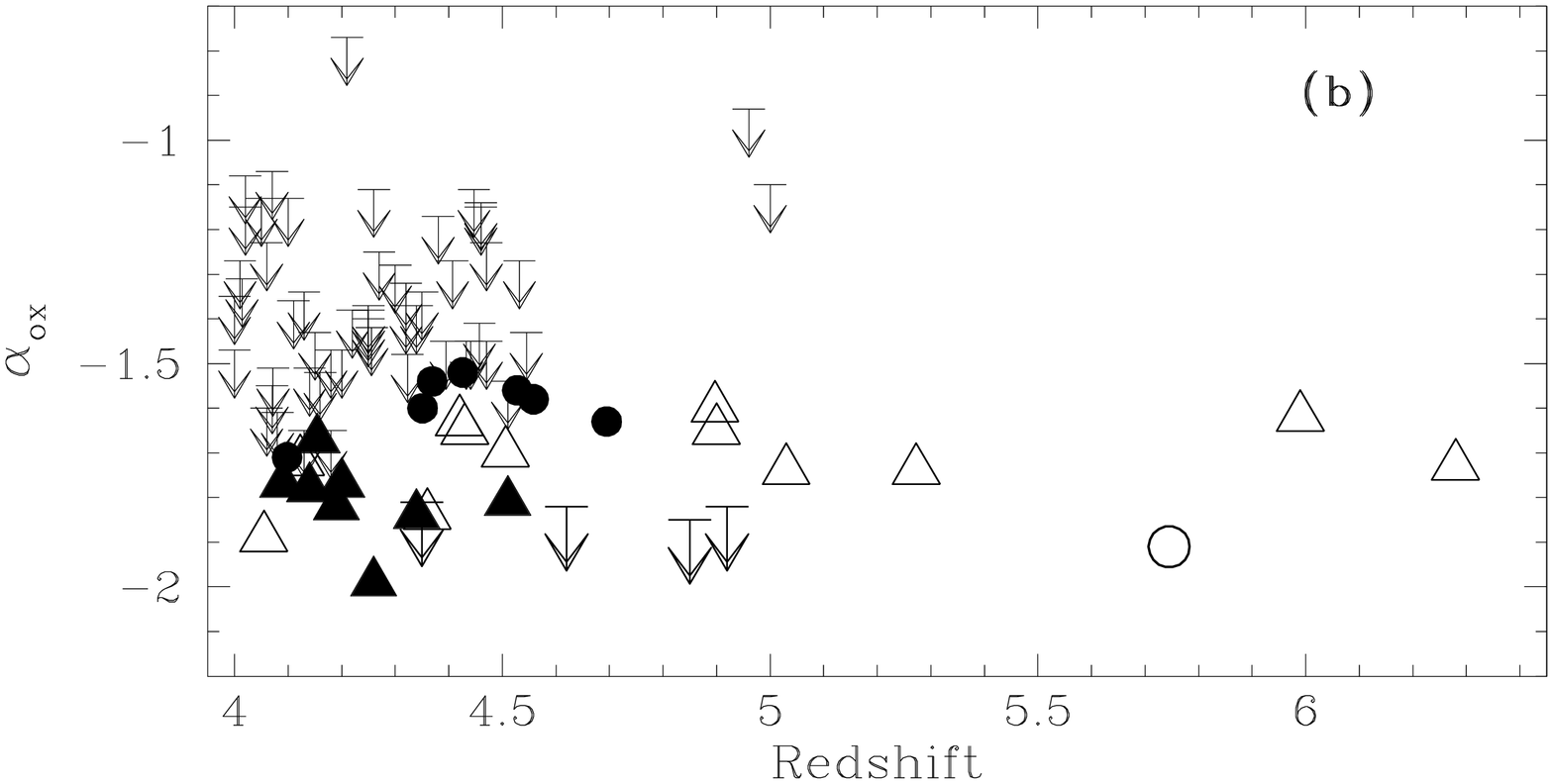}}
\vskip -1.1cm
\figcaption{\footnotesize 
(a) \aox\ versus $\log\ L_{2500~\rm \AA}$ and (b) redshift 
for $z\ge4$ optically selected RQQs. Symbols are the same as in Fig.~4. 
\label{fig5}}
\centerline{}
\centerline{}
%\end{figure}
%%%%%%%%%%%%%%%%%%
%%%	END of FIG.5
%%%%%%%%%%%%%%%%%%%%%%%%%%%%%%%%%%%%%%%%%%%%%%%%%%%%%%%%%%%%%%%%%%%%%%%%%%%
%
\noindent $z>4$ using the statistical methods available in {\sc ASURV} (see $\S$3.1) 
but found none. 
Given the small spread in UV luminosities and redshifts of the objects used in this analysis, 
we cannot address the question of 
\aox\ changes with rest-frame 2500~\AA\ luminosity or redshift 
effectively. However, the analysis of SDSS quasars in the redshift range $z\approx$~0.2--6.3 
using partial correlation analysis (Vignali, Brandt, \& Schneider 2002) shows 
that \aox\ is mainly dependent upon 2500~\AA\ luminosity.

\subsection{Joint spectral fit}

To provide tighter constraints on the average \hbox{X-ray} spectral properties of 
$z>4$ quasars, we have performed a joint spectral analysis of our target quasars. 
Source counts have been extracted from 2\arcsec\ radius circular apertures 
centered on the \hbox{X-ray} position of each quasar. The background has been taken 
from annuli centered on the sources, avoiding the presence of nearby faint \hbox{X-ray} sources. 
To account for the recently discovered quantum efficiency decay of 
ACIS\footnote{See http://cxc.harvard.edu/cal/Links/Acis/acis/Cal\_prods/qeDeg/index.html.} 
at low energies, possibly caused by  
molecular contamination of the ACIS filters, we have applied a time-dependent correction to the ACIS 
quantum efficiency generating a new ancillary response file (ARF) 
for each source.\footnote{See http://www.astro.psu.edu/users/chartas/xcontdir/xcont.html.} 
The quantum efficiency degradation is not severe at energies 
above $\approx$~0.7~keV ($\approx$~10\% at 1~keV), where most of the 
detected signal lies, but, if left uncorrected, it might affect the results at lower energies. 
In particular, the reduction in efficiency could produce unreliable column densities if not taken 
into account properly. 

Only two objects are characterized by low counting statistics ($\approx$~10 counts); 
all the others have $\approx$~20--70 counts in the 0.5--8~keV band. 
The sample does not seem to be biased by the presence 
of either a few high signal-to-noise ratio objects (see Table~2) or the 
RLQ PSS~0121$+$0347, whose X-ray photon index 
(derived from the band ratio analysis; see Table~2) is consistent with those of the other quasars. 
Spectral analysis was carried out with {\sc XSPEC}. 
The joint spectral fitting of the nine individual unbinned quasar spectra 
%(without background subtraction) 
was carried out using the Cash statistic (Cash 1979). 
The Cash statistic is particularly well suited to low-count sources 
(e.g.,\ Nousek \& Shue 1989), and, since it is applied to the unbinned data, it allows us to retain 
all the spectral information. Furthermore, 
we can apply to each source its own Galactic absorption and redshift. 
The resulting joint spectrum, although derived from only $\approx$~340 counts, 
is reasonably good due to the extremely low background of \chandra\ (there are only $\approx$~2.1 background 
counts in all source cells), and it allows spectral analysis in the $\approx$~2--30~keV rest-frame band. 
Joint fitting with a power-law model (leaving the normalizations free to vary) 
and Galactic absorption provides a good fit (as judged by the small data-to-model residuals 
and checked with 10000 Monte-Carlo simulations) with a 
photon index of $\Gamma$=1.98$\pm{0.16}$ (errors are at the 90\% confidence level; 
$\Delta\chi^{2}$=2.71, Avni 1976). 
This photon index is consistent with those obtained for lower-redshift samples of RQQs 
in the \hbox{2--10~keV} band (e.g., George et al. 2000; Mineo et al. 2000; Reeves \& Turner 2000) 
and with that of the BALQSO APM~08279$+$5255 at $z=3.91$ (Chartas et al. 2002; $\Gamma$=1.72$^{+0.06}_{-0.05}$). 
The photon index is also consistent with the average $\Gamma$ determined from the 
band ratios (see Table~2) using {\sc ASURV} ($\langle\Gamma\rangle$=2.07$\pm{0.13}$). 
The combined spectrum (used here only for presentation purposes), fitted with the power-law model 
described above, is shown in Fig.~6. 
Bechtold et al. (2002) reported an \hbox{X-ray} spectral flattening with redshift for quasars, 
with photon indices in the range $\Gamma\approx$~1.2--1.6 
($\langle\Gamma\rangle$=1.48$\pm{0.13}$) at \hbox{$z\approx$~3.7--6.3}. 
The present data, obtained for an optically brighter sample of quasars, 
are not consistent with such claims. 
%%%%%%%%%%%%%%
%%%	FIGURE 6: Joint X-ray spectrum of z>=4 QSOs + Confidence Contours for 
%		  an absorbed power-law model [3 cases]
%%%%%%%%%%%%%%%%%%
%\begin{figure}
\figurenum{6}
\vskip 1.2cm
\hglue -2.3cm
%\centerline{\includegraphics[height=0.25\textheight,angle=-90]{spectrum_res_cont.ps}}
\centerline{\includegraphics[width=2.5cm,angle=-90]{vignali.fig6.ps}}
\vskip 2.5cm
\figcaption{\footnotesize 
Combined quasar spectrum (used only for presentation purposes) fitted with 
the best-fit power-law model and Galactic absorption (see $\S$3.3 for details). 
The 68, 90, and 99\% confidence regions for the photon index and intrinsic column density 
obtained using the Cash statistic are shown in the insert. 
Data-to-model residuals are shown in the bottom panel (in units of $\sigma$). 
\label{fig6}}
%\end{figure}
\centerline{}
\centerline{}
%%%%%%%%%%%%%%%%%%
%%%	END of FIG. 6 
%%%%%%%%%%%%%%%%%%%%%%%%%%%%%%%%%%%%%%%%%%%%%%%%%%%%%%%%%%%%%%%%%%%%%%%%%%%

Furthermore, we have constrained the presence of neutral intrinsic absorption. 
In the absorption column we have assumed solar abundances, although there are indications of 
supersolar abundances of heavy elements in high-redshift quasar nuclei 
(e.g., Hamann \& Ferland 1999; Constantin et al. 2002). 
Doubling the abundances in the fit gives a reduction in the column density of a factor of 
$\approx$~2. 
There is no evidence for absorption in excess of the Galactic column density, as shown in the 
insert in Fig.~6, where the 68, 90, and 99\% confidence regions for the 
photon index and intrinsic column density are plotted. 
Although it is not possible to rule out substantial \hbox{X-ray} absorption in the individual spectrum 
of any one quasar, on average it appears that any neutral absorption has a column density of 
$N_{\rm H}\simlt8.8\times10^{21}$~cm$^{-2}$ (at 90\% confidence). 
Note that this average constraint is more generally applicable than if we had just 
a $\approx$~340-count spectrum of a single $z>4$ quasar, since any one quasar might always be exceptional. 
If our targets had intrinsic absorption of \hbox{$N_{\rm H}\approx$~(2--5)$\times10^{22}$~cm$^{-2}$}, 
as recently found in two RQQs at $z\approx$~2 using \asca\ data (Reeves \& Turner 2000), we would 
be able to detect it given the upper limit on the column density we derived. 
We have also investigated the possible presence of ionized absorption using the model {\sc ABSORI} in {\sc XSPEC}. 
While, of course, this is more difficult to constrain in general, our data do not provide any suggestion of ionized absorption. 
Furthermore, considering narrow iron K$\alpha$ lines in the rest-frame energy range 
6.4--6.97~keV, the typical upper limit on the equivalent width (EW) is $\approx$~420~eV.

\subsection{Companion objects}

Schwartz (2002b) tentatively associated an \hbox{X-ray} source 
23\arcsec\ from the $z=5.99$ quasar SDSS~1306$+$0356 with an \hbox{X-ray} jet, 
although neither the source nor SDSS~1306$+$0356 have any radio emission in the FIRST catalog 
(Becker et al. 1995). 
To date, no quasar with \hbox{X-ray} emission undoubtedly associated with jets is radio 
quiet, all being powerful radio sources (e.g., Harris \& Krawczynski 2002). 
Moreover, the jets are characterized by substantial radio emission (e.g., Siemiginowska et al. 2002; 
Brunetti et al. 2002 and references therein) even after correction for boosting. 
If the \hbox{X-ray} emission from jets is produced via the Compton scattering of cosmic 
microwave background (CMB) photons by relativistic electrons, then the 
enhancement of the CMB energy density with redshift compensates for the decrease of surface brightness 
(see Schwartz 2002a for details; also see Rees \& Setti 1968). 
Recently, Ivanov (2002) detected the \hbox{$I$-band} counterpart of the \hbox{X-ray} source 
23\arcsec\ from SDSS~1306$+$0356. The elongation of the \hbox{$I$-band} counterpart suggests the presence 
of a foreground galaxy, rather than UV/optical emission 
from a jet associated with SDSS~1306$+$0356, but no redshift for the counterpart 
has yet been measured.  

In the present sample of nine objects, 
only one quasar, PSS~0121$+$0347, is radio loud (see $\S$3). 
Nevertheless, we have searched for possible companions/jets for all the objects in our sample over a 
region of $\approx$~100\arcsec$\times$100\arcsec\ centered on the quasars 
(the same shown in the images of Fig.~2). 
At $z\approx$~4--4.5, this corresponds to a linear scale of $\approx$~320~kpc 
at each side of the quasar. 
We found some faint objects (about one per field); none of these is X-ray bright and close enough 
to our targets to contaminate their X-ray emission. 
Most of the serendipitous \hbox{X-ray} sources lack counterparts in the DPOSS images, and 
none of them has associated radio emission down to a 
1.4~GHz flux density of \hbox{$\approx$~0.5--1.5~mJy} (3$\sigma$). 

As an observational check, we compared the \hbox{0.5--2~keV} cumulative number counts of sources 
detected in the current high-redshift quasar fields 
with the number counts obtained from the \chandra\ Deep Field-North survey (Brandt et al. 2001a) 
and the \rosat\ deep survey of the Lockman Hole (Hasinger et al. 1998); 
the high-redshift quasars themselves, of course, were excluded. 
The surface density of \hbox{X-ray} sources surrounding the PSS quasars 
at the flux limit we reached \hbox{($\approx10^{-15}$ \cgs)} 
is $N(>S$)=1224$^{+612}_{-429}$~deg$^{-2}$, 
%$\log\ N=3.09^{+0.18}_{-0.19}$ deg$^{-2}$, 
consistent with the densities obtained in deep-field observations. 
Thus we find no evidence for an excess of companions/jets associated with our quasars.

\subsection{X-ray extension of the quasars}

To constrain the presence of either gravitational lensing 
(e.g., Barkana \& Loeb 2000; Wyithe \& Loeb 2002; Comerford, Haiman, \& Schaye 2002) 
or X-ray jets close to the sources (e.g., Schwartz 2002b), we 
performed an analysis of \hbox{X-ray} extension for all the quasars in the present sample. 
At $z\approx$~4--5, the angular separation between the images 
of a multiply imaged source should often be $\approx$~1--2\arcsec, depending on the adopted dark matter model 
(Barkana \& Loeb 2000), and \chandra\ has been able to detect and separate lensed images up to 
$z\approx$~3.9 (Chartas et al. 2002). 
For each quasar, we created a point spread function (PSF) at $\approx$~1.5 keV 
at the source position (using the tool {\sc MKPSF}) 
normalized by the effective number of source counts. 
Then we compared the radial profile of this PSF with that of the source, finding good agreement in 
the radial counts distribution. 
For one quasar (PSS~0926$+$3055) we performed a further check given the presence in the soft 
band of three counts ``pointing'' in the North-West direction within a radius of 3\arcsec\ of the 
source, although there is no indication of extension in the optical image. 
We deconvolved the \hbox{X-ray} source image with the PSF and found that 
the counts distribution still seems to be in agreement with the source being point-like. 
However, further investigation with a longer \chandra\ observation is needed to completely rule 
out the possibility of PSS~0926$+$3055 being slightly extended.

\subsection{X-ray variability of the quasars}

Although the present observations are short (the observation length in the rest frame 
is $\approx$~15 minutes), we searched for \hbox{X-ray} variability of our targets. 
There are tentative reports that quasar \hbox{X-ray} variability 
increases with redshift (e.g., Manners, Almaini, \& Lawrence 2002). 
We analyzed the photon arrival times in the \hbox{0.5--8~keV} band 
using the Kolmogorov-Smirnov (KS) test. 
No significant variability was detected (the KS probabilities that the sources are variable 
range from $\approx$~6\% to 83\%). 
Our constraints on the amplitude of variability depend upon the number of counts detected 
as well as the putative variability timescale. 
Systematic variability sustained over a period of $\approx$~5--6~ks (in the observed frame) 
must be less than $\approx$~30--90\%. 
%
%%% METHOD %%%%
% Brightest source: 68 counts FB; divide the observation by 2 to have enough counts/interval
% ==> 34 counts/interval; sqrt(34)=5.8 ==> 5.8/34=0.17 multiplied by 2 (because the 
% variability data points must be clearly seen (outside 1 sigma) ==>> 0.34=34%
% For the faintest X-ray source: 10 FB counts => 2.2/5=0.45x2=0.90=90%
%%%%%%%%%%%%%%%

\section{Optical analysis}

Following the method described in $\S$2.2, 
two quasars, PSS~0926$+$3055 and PSS~0955$+$5940, have been found to vary in the optical band 
on a timescale of $\approx$~6 years (in the observed 
%
%%%%%%%%%%%%%%%%%%%%%%%%%%%%%%%%%%%%%%%%%%%%%%%%%%%%%%%%%%%%%%%%%%%%%%%%%%%
%%%	TABLE 4: Optical Emission-Line measurements
%%%%%%%%%%%%%%%%%%
\end{multicols}
\begin{deluxetable}{lcrccccccrccccrcccccrcc}
\rotate
\tablecolumns{23}
\tabletypesize{\scriptsize}
\tablewidth{0pt}
\tablecaption{Emission-Line Measurements}
\tablehead
{
\colhead{} & \multicolumn{2}{c}{Ly$\alpha$ (1216~\AA)} & 
\colhead{} & \multicolumn{3}{c}{N~V (1240~\AA)} & 
\colhead{} & \multicolumn{4}{c}{O~I (1304~\AA)} & 
\colhead{} & \multicolumn{4}{c}{Si~IV-O~IV] (1394-1402~\AA)} & 
\colhead{} & \multicolumn{4}{c}{C~IV (1549~\AA)} & \colhead{} \\
\cline{2-3} \cline{5-7} \cline{9-12} \cline{14-17} \cline{19-22} \\
\colhead{Object Name} & 
\colhead{$\lambda$\tablenotemark{a}} & \colhead{EW\tablenotemark{b}} & 
\colhead{} & 
\colhead{$\lambda$\tablenotemark{a}} & \colhead{EW\tablenotemark{b}} & \colhead{$z$} & 
\colhead{} & 
\colhead{$\lambda$\tablenotemark{a}} & \colhead{EW\tablenotemark{b}} & 
\colhead{FWHM\tablenotemark{c}} & \colhead{$z$} & \colhead{} & 
\colhead{$\lambda$\tablenotemark{a}} & \colhead{EW\tablenotemark{b}} & 
\colhead{FWHM\tablenotemark{c}} & \colhead{$z$} & \colhead{} & 
\colhead{$\lambda$\tablenotemark{a}} & \colhead{EW\tablenotemark{b}} & 
\colhead{FWHM\tablenotemark{c}} & \colhead{$z$} & \colhead{$\langle z\rangle$} 
}
\startdata
PSS~0121$+$0347 & 
6232.3 & 81.1 & & 
\nodata & \nodata & \nodata & & 
6674.4 & 4.8 & 1020 & 4.112 & &
\nodata & \nodata & \nodata & \nodata & & 
7963.6 & 33.5 & 3810 & 4.142 & 
4.126 \\
PSS~0133$+$0400 & 
6266.1 & 25.9 & & 
6391.7 & 2.3 & 4.152 & &
6734.8 & 10.3 & 6270 & 4.158 & &
7218.4 & 5.3 & 2550 & 4.150 & &
7999.7 & 32.6 & 7370 & 4.165 &
4.155 \\
PSS~0209$+$0517 & 
6326.1 & 33.5 & & 
\nodata & \nodata & \nodata & & 
\nodata & \nodata & \nodata & \nodata & & 
7216.8 & 1.1 & 3350 & 4.149 & & 
7945.1 & 12.0 & 12940 & 4.130 & 
4.143 \\
PSS~0926$+$3055 & 
6309.0 & 44.0 & & 
\nodata & \nodata & \nodata & &  
6802.6 & 1.3 & 1460 & 4.210 & & 
7261.6 & 8.2 & 6620 & 4.181 & &
8028.1 & 18.5 & 5880 & 4.184 & 
4.191 \\
PSS~0955$+$5940 & 
6503.5 & 285.0 & &
\nodata & \nodata & \nodata & & 
6979.2 & 3.9 & 2270 & 4.345 & & 
7471.1 & 4.0 & 2730 & 4.330 & &
8270.0 & 21.6 & 4220 & 4.340 & 
4.339 \\
PSS~0957$+$3308 & 
6386.9 & 22.1 & &
\nodata & \nodata & \nodata & & 
6829.5 & 1.2 & 2150 & 4.231 & &
7267.2 & 9.4 & 6340 & 4.185 & & 
8020.7 & 12.1 & 9500 & 4.179 & 
4.198 \\
PSS~1326$+$0743 & 
6223.6 & 26.2 & &
6306.3 & 4.2 & 4.083 & & 
6670.6 & 1.6 & 2260 & 4.109 & & 
7122.7 & 10.0 & 7130 & 4.082 & &
7905.4 & 21.2 & 10920 & 4.105 & 
4.089 \\
PSS~1347$+$4956 & 
6760.6 & 8.7 & &
\nodata & \nodata & \nodata & &  
\nodata & \nodata & \nodata & \nodata & & 
%7422.7 & 0.7 & 2130 & &
7708.4 & 4.1 & 5630 & 4.500 & &
8535.3 & 15.9 & 7670 & 4.511 & 
4.506 \\ 
PSS~1443$+$5856 & 
6465.0 & 69.5 & &
\nodata & \nodata & \nodata & & 
\nodata & \nodata & \nodata & \nodata & & 
7369.2 & 9.9 & 6760 & 4.258 & &
8161.8 & 19.0 & 7980 & 4.270 & 
4.264 \\
\tableline
\enddata
\tablecomments{The redshift for each quasar was determined by averaging the emission lines 
(excluding Ly$\alpha$ for all objects). The last column reports the weighted average for the redshift. 
The FWHMs of Ly$\alpha$ and N~V are not reported because of the uncertainties 
in their measurements (see, e.g., Schneider et al. 1991).}
\tablenotetext{a}{Observed-frame wavelength (\AA).}
\tablenotetext{b}{Rest-frame equivalent width (\AA).}
\tablenotetext{c}{FWHM of the line (km s$^{-1}$), corrected for the instrumental resolution.}
\label{tab4}
\end{deluxetable}
\begin{multicols}{2}
%%%%%%%%%%%%%%%%%%%%%
%%%	End of Table 4
%%%%%%%%%%%%%%%%%%%%%%%%%%%%%%%%%%%%%%%%%%%%%%%%%%%%%%%%%%%%%%%%%%%%%%%%%%%
%

\noindent 
frame). 
The HET $R$-band magnitude for the former is 0.3 mag brighter than the value reported in the Djorgovski list (taking 
into account the different filter transmission and bandpass; see Footnote~4), while the latter appears 0.5 mag fainter. 
These are the only objects for which significant long-term optical variability has been found. 

The principal emission-line measurements are presented in Table~4. 
The centroids of the emission lines were determined with Gaussian fits. 
Redshifts have been derived from HET spectra averaging the emission-line redshifts 
(excluding Ly$\alpha$, whose measurements are affected by many uncertainties; 
see, e.g., Schneider, Schmidt, \& Gunn 1991). 
Typical uncertainties are $\approx$~0.01 for non-BALQSO redshift measurements, and 
$\approx$~10--15\% and $\approx$~5--10\% for FWHM and EW measurements, respectively. 
%%
%It must be kept in mind that the Ly$\alpha$ line measurements 
%may be affected by systematic errors due to the proximity of the Ly$\alpha$ forest. 
%%
We have compared the \ion{C}{4} emission-line equivalent widths of our objects with those derived for 
the Constantin et al. (2002) sample of optically luminous $z>4$ PSS and BRI quasars; we found 
good agreement. 
One of our quasars, PSS~1443$+$5856, shows broad \ion{C}{4} and \ion{Si}{4} absorption features 
in its optical spectrum  (see Fig.~3) with a blueshift of \hbox{$\approx$~10000~km s$^{-1}$} 
with respect to the systemic redshift of the quasar. 
The \ion{C}{4} absorption-line EW is $\approx$~14~\AA\ (rest frame), which is close to the value expected from the 
\hbox{\aox--\ion{C}{4} EW} anti-correlation found for low-redshift quasars (see Fig.~4 of Brandt et al. 2000). 
The likely existence of this anti-correlation at $z>4$ has already been verified 
(e.g., Maiolino et al. 2001; Goodrich et al. 2001; also see V01). 
Given the correlation between \ab1450\ and soft \hbox{X-ray} flux shown in Fig.~4, the \hbox{X-ray} detection of 
PSS~1443$+$5856 is not surprising. Since this quasar is 1.5--1.7 magnitudes brighter than the previously 
\hbox{X-ray} undetected $z>4$ BALQSOs (V01), a higher \hbox{X-ray} flux is expected.

\section{Summary and implications}

We have presented the \chandra\ X-ray detections of nine high-redshift \hbox{($z\approx$~4.1--4.5)} 
quasars. These quasars were targeted because they are among the most optically luminous objects in the Universe 
and are likely to comprise a significant fraction of the most luminous $z>4$ optically selected objects. 
Their SEDs are characterized by steeper (more negative) optical-to-X-ray spectral indices 
($\langle\alpha_{\rm ox}\rangle$=$-$1.81$\pm{0.03}$) compared to samples of optically selected quasars at 
low or intermediate redshift. Given their large mean 2500~\AA\ luminosity density 
($\approx$~1.6$\times10^{32}$~\lumh), their steep \aox\ values may be explained in the context of the \aox--ultraviolet 
luminosity anti-correlation found in other work (e.g., Avni et al. 1995; Vignali et al. 2002). 
We also confirm the presence of a significant correlation between the optical magnitude and 
soft X-ray flux previously found by V01 for a more heterogeneous sample of $z>4$ quasars. 
Joint spectral fitting in the \hbox{$\approx$~2--30~keV} rest-frame band indicates that 
optically selected, luminous $z>4$ quasars are characterized, on average, by a power-law model with a 
photon index of $\approx$~2.0$\pm{0.2}$, similar to the majority of 
lower-redshift quasars. There is no evidence for intrinsic absorption 
($N_{\rm H}\simlt8.8\times10^{21}$~cm$^{-2}$ at the 90\% confidence level).  

We also have presented near-simultaneous optical spectra obtained with the HET. 
Optical spectra for seven of these quasars are published here for the first time. 
One of the quasars, PSS~1443$+$5856, shows broad absorption features in \ion{C}{4} and \ion{Si}{4}. 

From a more general perspective, our results show that 
$z\approx$~4.1--4.5 quasars and local quasars have reasonably 
similar \hbox{X-ray} and broad-band spectra (once luminosity effects 
are taken into account). The small-scale \hbox{X-ray} emission regions 
of quasars do not appear to be sensitive to the substantial 
large-scale environmental differences between $z\approx 4$ 
and $z\approx 0$. The lack of any strong \hbox{X-ray} spectral changes 
is particularly relevant since \hbox{X-ray} emission originates 
from the compact region where accretion, and thus black hole 
growth, occurs. Our data do not provide any hints for different 
accretion/growth mechanisms (via, e.g., accretion-disk instabilities 
or ``trapping radius'' effects) between the most luminous 
high-redshift quasars and local quasars. Furthermore, our 
data suggest that (1) minimum black hole masses for high-redshift 
quasars can be estimated reasonably well via the Eddington limit 
by adopting the bolometric correction used for comparably 
luminous low-redshift quasars, and (2) calculations of the 
effects of early quasars upon the intergalactic medium can, at 
least to first order, adopt the spectral energy distributions 
of comparably luminous low-redshift quasars. Finally, our data 
support the dogma that \hbox{X-ray} emission is a universal property 
of quasars at all redshifts. Quasars at even higher redshifts 
($z\approx$~5--15) will likely be luminous \hbox{X-ray} sources that 
can be detected in future X-ray surveys.
%%%%%%%%%%%%%%
%%%	FIGURE 7: Constellation-X simulation of 40 ks power-law (Gamma=2.0)+ Nh=5e21 + 
%		  Gaussian (sigma=50 eV RF; EW=32 eV RF) model
%%%%%%%%%%%%%%%%%%
%\begin{figure}
\figurenum{7}
\vskip -1.2cm 
%\centerline{\includegraphics[width=0.35\textwidth,angle=-90]{spectrum_40ks_both.ps}}
\centerline{\includegraphics[width=6cm,angle=-90]{vignali.fig7a.eps}}
\vglue0.4cm
%\centerline{\includegraphics[width=0.35\textwidth,angle=-90]{ionized_narrow.ps}}
\centerline{\includegraphics[width=6cm,angle=-90]{vignali.fig7b.eps}}
\vglue 0.4cm
%\centerline{\includegraphics[width=0.35\textwidth,angle=-90]{nh_2e21_cont.ps}}
\centerline{\includegraphics[width=6cm,angle=-90]{vignali.fig7c.eps}}
\vskip 0.25cm
\figcaption{\footnotesize 
(a) Simulated 40~ks spectrum and (b) 68, 90, and 99\% confidence regions for the He-like iron K$\alpha$ 
emission-line energy and intensity using the calorimeter and the grating onboard \conx. 
We have used the observed \chandra\ \hbox{0.5--2~keV} flux of $\approx$~2.8\ $\times\ 10^{-14}$~\cgs and 
assumed a $\Gamma=2.0$ power-law spectrum with Galactic absorption. 
We also included a narrow iron K$\alpha$ emission line 
with a rest-frame equivalent width of 70~eV (see $\S$5 for details). 
(c) 68, 90, and 99\% confidence regions for intrinsic column density and photon index. 
The ``X'' marks the input parameters of the simulation, while the ``$+$'' indicates the best-fit result. 
\label{fig7}}
%\end{figure}
\centerline{}
\centerline{}
%%%%%%%%%%%%%%%%%%
%%%	END of FIG. 7 
%%%%%%%%%%%%%%%%%%%%%%%%%%%%%%%%%%%%%%%%%%%%%%%%%%%%%%%%%%%%%%%%%%%%%%%%%%%

\section{Future work}

We have recently been awarded 11 additional \chandra\ observations of $z>4$ optically luminous quasars. 
These observations will provide tighter constraints 
on the \hbox{X-ray} properties of high-redshift quasars, thus 
allowing for more feasible planning of \hbox{X-ray} observations with next-generation, large-area 
\hbox{X-ray} telescopes. 
In this regard, 
we have simulated a 40~ks \conx\footnote{See http://constellation.gsfc.nasa.gov/docs/science/matrices.html.} 
\ spectrum (with the calorimeter and the first-order grating; see Fig.~7a) 
of the \hbox{X-ray} brightest RQQ of our sample, PSS~0926$+$3055 at $z=4.19$, 
assuming a power-law spectrum with $\Gamma$=2.0 and Galactic absorption. 
A narrow ($\sigma$=10~eV), He-like (at 6.7~keV) iron K$\alpha$ emission line 
with a rest-frame equivalent width of $\approx$~70~eV has also been included in the model. 
At high \hbox{X-ray} luminosities \hbox{($L_{\rm 2-10~keV}$$>$10$^{44-45}$ \lum)} 
the iron K$\alpha$ line is expected to be ionized and fairly weak, with most of the line flux coming 
from its core (the ``X-ray Baldwin effect'' discussed in Nandra et al. 1997; see also Iwasawa \& Taniguchi 1993). 
The 68, 90, and 99\% confidence regions for the line energy and intensity are plotted in Fig.~7b. 
\conx\ would gather $\approx$~11300 counts ($\approx$~8500 from the calorimeter and 
$\approx$~2800 from the first-order grating) 
in the observed \hbox{0.25--10~keV} band. 
These would ensure a good accuracy in the measurements of the \hbox{X-ray} parameters; in particular, 
the photon index, line energy, and line EW would be measured with accuracies of 
$\approx$~1.2\%, 0.1\%, and 29\%, respectively. 
We are able to detect ionized iron K$\alpha$ lines down to rest-frame EWs of $\approx$50~eV. 
Due to its low-energy coverage, \conx\ can also place significant constraints on the intrinsic column density 
down to $N_{\rm H}\approx2\times10^{21}$~cm$^{-2}$ (see Fig.~7c).

\acknowledgments

We gratefully acknowledge the financial support of NASA grant NAS~8-01128 (GPG, principal investigator), 
NASA LTSA grant NAG5-8107 (CV), 
\chandra\ X-ray Center grant DD1-2012X (CV, WNB, DPS), and NSF grant AST99-00703 (DPS). 
We thank D. Alexander, G. Brunetti, G. Chartas, E. Feigelson, S. Gallagher, 
S. Park, L. Pozzetti, K. Weaver, and N. White for useful discussions, 
F. Bauer for help with IDL codes, M. Brusa, M. Mignoli, and G. Richards for help 
with the reduction of the optical spectra, and the referee for useful comments. 
We thank S.~G. Djorgovski and the members of the DPOSS team for making their quasars 
available to the community. 
CV also acknowledges partial support from the Italian Space Agency, under the contract 
ASI 00/IR/103/AS, and from the Ministry for University and Research (MURST) 
under grant Cofin-00-02-36. 

The Hobby-Eberly Telescope (HET) is a joint project of the University of Texas 
at Austin, the Pennsylvania State University, Stanford University, 
Ludwig-Maximillians-Universit\"at M\"unchen, and Georg-August-Universit\"at G\"ottingen. 
The HET is named in honor of its principal benefactors, 
William P. Hobby and Robert E. Eberly. The Marcario Low-Resolution 
Spectrograph is named for Mike Marcario of High Lonesome Optics, who 
fabricated several optics for the instrument but died before its completion; 
it is a joint project of the Hobby-Eberly Telescope partnership and the 
Instituto de Astronom\'{\i}a de la Universidad Nacional Aut\'onoma de M\'exico.

\clearpage

\appendix
% OLD NUMBERS: +2 RQQs + 1 RLQ observed by Chandra: 48 Upper Limits (35 RQQs, 6 RLQs, 5 loose RQQs and 2 without radio coverage)

\section{New \rosat\ upper limits for $z\ge4$ quasars and a 
tentative detection of GB~1713$+$2148}

The Djorgovski list of $z\ge4$ quasars (see Footnote~4) has been cross correlated with \rosat\ PSPC and HRI 
archival observations. In the \hbox{0.5--2~keV} energy band, 
we found 45 new upper limits [34 RQQs, five RLQs, four quasars with loose radio loudness upper limits ($R\simlt15$), 
and two quasars without FIRST/NVSS coverage] and a tentative 
$\approx$~3$\sigma$ detection for the blazar GB~1713$+$2148 (in an HRI observation). 
The 3$\sigma$ upper limits (reported in Table~A1) have been 
derived in a similar way to those shown in Table~4 of V01 
using the {\sc SOSTA} command in the {\sc XIMAGE} package (Giommi et al. 1992). 
For the 3$\sigma$ detection of GB~1713$+$2148, 
we cross checked the results obtained with the maximum likelihood method available with the 
\hbox{{\sc MIDAS/EXSAS}} package (Zimmermann et al. 1998) with the {\sc XIMAGE} results. 
GB~1713$+$2148 is unusually \hbox{X-ray} faint compared to the other blazars, 
despite its large radio flux density ($\approx$~450~mJy in the NVSS; $R\approx$~24400). 

For SDSS quasars, the \ab1450\ magnitudes have been derived from the $i$-band filter 
using either the empirical relationship 
\ab1450=$i-0.2$
or the method described in Vignali et al. (2002) 
for the SDSS quasars in the Early Data Release (Schneider et al. 2002). 
For the other quasars, the \ab1450\ magnitudes have been 
derived either from the $R$-band magnitude using the empirical relationship 
\ab1450=$R-$0.684\ $z+3.10$
or from the $r$-band magnitude using 
\ab1450=$r-$0.684\ $z+2.75$.
Unless otherwise stated, the $R$- and $r$-band magnitudes 
have been obtained from the Djorgovski list (see Footnote~4). 
All of the above relationships provide reliable \ab1450\ estimates (within $\approx$~0.1--0.2 magnitudes) 
in the redshift range under consideration. 
%%%
%%%%%%%%%%%%%%%%%%%%%%%%%%%%%%%%%%%%%%%%%%%%%%%%%%%%%%%%%%%%%%%%%%%%%%%%%%%
%%%	APPENDIX (A1): ROSAT Optical + X-ray and radio properties: NEW analysis
%%%%%%%%%%%%%%%%%%
\begin{deluxetable}{lccccc}
%\rotate
\tablenum{A1}
\tablecolumns{6}
\tabletypesize{\footnotesize}
\tablewidth{0pt}
\tablecaption{Properties of $z\ge4$ Quasars in \rosat\ Fields}
\tablehead{ 
\colhead{Object} & \colhead{$z$} & \colhead{$AB_{1450(1+z)}$} & \colhead{$M_B$} & \colhead{$f_{\rm x}$\tablenotemark{a}} & 
\colhead{$\alpha_{\rm ox}$} 
}
\startdata
%%%%
\cutinhead{\rosat\ RQQs}
PSS~0014$+$3032    & 4.47 & 18.5 & $-$28.6 & $<4.44$ & $<-1.45$ \\
BRI~0046$-$2458    & 4.15 & 18.9 & $-$28.1 & $<3.97$ & $<-1.43$ \\
SDSS~0127$-$0045   & 4.06 & 17.9 & $-$29.0 & $<3.53$ & $<-1.60$ \\
PSS~0134$+$3307    & 4.53 & 18.5 & $-$28.6 & $<13.3$ & $<-1.27$ \\
PSS~0223$+$2325    & 4.27 & 18.8 & $-$28.2 & $<12.3$ & $<-1.25$ \\
SDSS~0239$-$0810   & 4.00 & 19.4 & $-$27.4 & $<1.90$ & $<-1.47$ \\
SDSS~0252+0031     & 4.10 & 19.7 & $-$27.2 & $<11.8$ & $<-1.13$ \\
BR~0300$-$0207     & 4.25 & 18.4 & $-$28.6 & $<8.46$ & $<-1.37$ \\ 
BR~0353$-$3820     & 4.55 & 17.6 & $-$29.5 & $<11.7$ & $<-1.43$ \\ 
FIRST~0747$+$2739\tablenotemark{b}  
                   & 4.11 & 18.2 & $-$28.7 & $<11.3$ & $<-1.36$ \\
PSS~0808$+$5215    & 4.44 & 18.5 & $-$28.5 & $<4.58$ & $<-1.45$ \\
SDSS~0810$+$4603   & 4.07 & 18.2 & $-$28.7 & $<7.03$ & $<-1.55$ \\
SDSS~0832$+$5303   & 4.02 & 19.6 & $-$27.3 & $<22.0$ & $<-1.08$ \\
PSS~0852$+$5045    & 4.20 & 18.0 & $-$29.0 & $<7.24$ & $<-1.47$ \\
SDSS~0904$+$5350   & 4.25 & 19.1 & $-$27.9 & $<3.90$ & $<-1.40$ \\
%PSS~0957$+$3308\tablenotemark{b}
%                   & 4.20 & 18.2 & $-$28.7 & $<9.81$ & $<-1.38$ \\
PSS~1026$+$3828    & 4.18 & 18.8 & $-$28.1 & $<3.20$ & $<-1.47$ \\
SDSS~1034$-$0027   & 4.38 & 19.8 & $-$27.2 & $<7.48$ & $<-1.17$ \\
SDSS~1040$-$0015   & 4.32 & 18.7 & $-$28.3 & $<8.28$ & $<-1.32$ \\
SDSS~1048$-$0028   & 4.00 & 19.2 & $-$27.7 & $<4.66$ & $<-1.35$ \\ 
SDSS~1052$-$0006   & 4.13 & 19.6 & $-$27.3 & $<3.33$ & $<-1.34$ \\
SDSS~1056$+$0032   & 4.02 & 19.8 & $-$27.1 & $<9.52$ & $<-1.15$ \\
SDSS~1059$+$0104   & 4.06 & 19.4 & $-$27.6 & $<8.61$ & $<-1.23$ \\ 
PSS~1140$+$6205    & 4.51 & 18.4 & $-$28.7 & $<2.86$ & $<-1.54$ \\
PSS~1226$+$0950    & 4.34 & 18.6 & $-$28.4 & $<7.04$ & $<-1.37$ \\
PSS~1248$+$3110    & 4.32 & 19.1 & $-$27.9 & $<4.60$ & $<-1.37$ \\  
PSS~1315$+$2924    & 4.18 & 18.4 & $-$28.6 & $<1.66$ & $<-1.65$ \\
PSS~1339$+$5154    & 4.08 & 19.0 & $-$27.9 & $<1.21$ & $<-1.61$ \\
%PSS~1347$+$4956\tablenotemark{b}
%                   & 4.51 & 17.4 & $-$29.7 & $<9.89$ & $<-1.49$ \\
PSS~1401$+$4111    & 4.01 & 18.6 & $-$28.2 & $<13.2$ & $<-1.27$ \\
PSS~1418$+$4449    & 4.32 & 18.5 & $-$28.5 & $<3.69$ & $<-1.48$ \\
PSS~1430$+$2828    & 4.30 & 19.5 & $-$27.5 & $<5.38$ & $<-1.28$ \\
PSS~1543$+$3417    & 4.41 & 18.5 & $-$28.5 & $<13.7$ & $<-1.27$ \\
PSS~1555$+$2003    & 4.22 & 19.1 & $-$27.8 & $<4.26$ & $<-1.38$ \\
PSS~1745$+$6846    & 4.13 & 19.1 & $-$27.9 & $<0.87$ & $<-1.65$ \\
PSS~2155$+$1358\tablenotemark{c}   
                   & 4.26 & 18.2 & $-$28.8 & $<7.73$ & $<-1.42$ \\
\cutinhead{\rosat\ RLQs}
%
%PSS~0121$+$0347\tablenotemark{b} 
%                   & 4.13 & 18.5 & $-$28.4 & $<6.67$ & $<-1.40$ \\
SDSS~0300$+$0032   & 4.19 & 19.9 & $-$27.0 & $<5.19$ & $<-1.22$ \\
PSS~0439$-$0207    & 4.40 & 18.8 & $-$28.2 & $<5.55$ & $<-1.37$ \\
SDSS~0913$+$5919\tablenotemark{d}
%{c}
                   & 5.11 & 20.3 & $-$27.0 & $<8.70$ & $<-1.05$ \\
CLASS~0915$+$068   & 4.19 & 18.9 & $-$28.0 & $<7.08$ & $<-1.32$ \\
CLASS~1322$+$116   & 4.40 & 18.9 & $-$28.1 & $<7.47$ & $<-1.31$ \\
\cutinhead{\rosat\ quasars with loose radio loudness upper limits}
SDSS~0338$-$RD657\tablenotemark{e}  
                   & 4.96 & 21.1 & $-$26.1 & $<8.63$ & $<-0.93$ \\
SDSS~0405$+$0059   & 4.05 & 19.8 & $-$27.1 & $<10.0$ & $<-1.13$ \\ 
SDSS~1708$+$6022   & 4.35 & 19.8 & $-$27.2 & $<2.60$ & $<-1.34$ \\
SDSS~1710$+$5923   & 4.47 & 19.7 & $-$27.4 & $<6.00$ & $<-1.23$ \\
\cutinhead{\rosat\ quasars without FIRST/NVSS coverage}
BR~0413$-$4405\tablenotemark{c}      
		   & 4.07 & 19.2 & $-$27.7 & $<25.1$ & $<-1.07$ \\
BR~0418$-$5723\tablenotemark{c}      
                   & 4.46 & 18.8 & $-$28.3 & $<21.6$ & $<-1.15$ \\
\cutinhead{\rosat\ Blazar}
GB~1713$+$2148 & 4.01 & 21.4 & $-$25.5 & 2.81 & $-$1.11 \\
%%%%
\tableline
\enddata
\tablecomments{The division between RQQs and RLQs is based on the radio-loudness parameter (defined 
in $\S$3).}
\tablenotetext{a}{Galactic absorption-corrected flux in the observed 0.5--2 keV band in units 
of $10^{-14}$ erg cm$^{-2}$ s$^{-1}$.} 
%\tablenotetext{b}{Also observed by \chandra\ and presented in this paper; see Table~3.}
\tablenotetext{b}{\ab1450\ magnitude derived from the $R$-band magnitude published in 
Richards et al. (2002); also the redshift is from Richards et al. (2002).}
\tablenotetext{c}{\ab1450\ magnitude derived from the $R$-band magnitude published in Peroux et al. 
(2001); also the redshift for BR~0418$-$5723 has been obtained from Peroux et al. (2001).}
\tablenotetext{d}{Also observed by \chandra\ (C. Vignali et al., in preparation).}
\tablenotetext{e}{Also observed by \xmm\ (C. Vignali et al., in preparation).}
\label{A1}
\end{deluxetable}
%%%%%%%%%%%%%%%%%%%%%
%%%	End of Table 4
%%%%%%%%%%%%%%%%%%%%%%%%%%%%%%%%%%%%%%%%%%%%%%%%%%%%%%%%%%%%%%%%%%%%%%%%%%%

\end{document}